\documentclass{article}

\usepackage{arxiv}

\usepackage[utf8]{inputenc} 
\usepackage[T1]{fontenc}    
\usepackage{hyperref}       
\usepackage{url}            
\usepackage{booktabs}       
\usepackage{amsfonts}       
\usepackage{nicefrac}       
\usepackage{microtype}      
\usepackage{lipsum}		
\usepackage{graphicx}
\usepackage{natbib}
\usepackage{doi}
\usepackage{amsthm}
\usepackage{amsmath}
\usepackage{pifont}
\usepackage{multirow}
\usepackage{graphicx,subfigure}
\usepackage{todonotes}
\usepackage{diagbox}
\newtheorem{definition}{Definition}

\newcommand{\xmark}{\text{\ding{55}}}
\newcommand{\cmark}{\text{\ding{51}}}

\title{Cardinality Estimation over Knowledge Graphs with
Embeddings and Graph Neural Networks}

\author{ \href{https://orcid.org/0009-0009-7957-603X}{\includegraphics[scale=0.06]{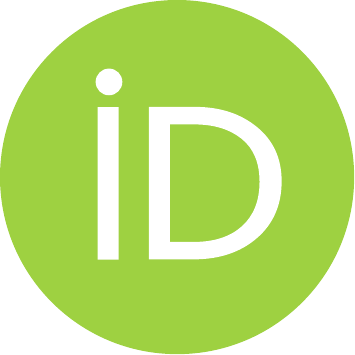}\hspace{1mm}Tim Schwabe\thanks{Work partially conducted at Ruhr University Bochum.}}\\
	Technical University of Munich\\
	TUM School of Computation, Information and Technology, Germany\\
	\And
	\href{https://orcid.org/0000-0002-1209-2868}{\includegraphics[scale=0.06]{orcid.pdf}\hspace{1mm}Maribel Acosta} \\
	Technical University of Munich\\
	TUM School of Computation, Information and Technology, Germany\\
}

\renewcommand{\shorttitle}{\small{Cardinality Estimation over Knowledge Graphs with
Embeddings and Graph Neural Networks}}

\hypersetup{
pdftitle={A template for the arxiv style},
pdfsubject={q-bio.NC, q-bio.QM},
pdfauthor={David S.~Hippocampus, Elias D.~Striatum},
pdfkeywords={First keyword, Second keyword, More},
}

\begin{document}
\maketitle

\begin{abstract}
Cardinality Estimation over Knowledge Graphs (KG) is crucial for query optimization, yet remains a challenging task due to the semi-structured nature and complex correlations of data in typical KGs. 
In this work, we propose \textit{GNCE}, a novel approach that leverages knowledge graph embeddings and Graph Neural Networks (GNN) to accurately predict the cardinality of conjunctive queries over KGs. 
\textit{GNCE} first creates semantically meaningful embeddings for all entities in the KG, which are then used to learn a representation of a query using a GNN to estimate the cardinality of the query. 
We evaluate \textit{GNCE} on several KGs in terms of q-Error and demonstrate that it outperforms state-of-the-art approaches based on sampling, summaries, and (machine) learning in terms of estimation accuracy while also having a low execution time and few parameters.  Additionally, we show that \textit{GNCE} performs similarly well on real-world queries and can inductively generalize to unseen entities, making it suitable for use in dynamic query processing scenarios. 
Our proposed approach has the potential to significantly improve query optimization and related applications that rely on accurate cardinality estimates of conjunctive queries.
\end{abstract}

\keywords{Cardinality Estimation \and Knowledge Graphs \and Machine Learning}

\section{Introduction}
Knowledge Graphs (KGs) allow for representing interconnected data, using semi-structured data models like the Resource Description Framework (RDF)~\cite{rdf}.   
In KGs, statements are represented in the form of a graph, where labeled nodes typically correspond to entities, and directed, labeled edges are connections between the nodes. 
Nowadays, KGs are fueling diverse applications like question-answering, recommender systems, and semantic databases~\cite{natasha}.

Similar to other database technologies, the data encoded in a KG can be queried using a structured  
query language like SPARQL~\cite{sparql}. 
In particular, conjunctive queries over KGs can be formulated also as graphs, where nodes and edges can be either \textit{constants} (i.e., values matched over the KG) or \textit{variables} (i.e., placeholders). 
The computation of answers for a given conjunctive query is carried out as a graph pattern-matching process, where the query cardinality corresponds to the number of subgraphs in the KG that are homomorphic to the query graph. 
To efficiently execute a (conjunctive) query over a KG, query engines implement optimization techniques that rely on cardinality (alt. selectivity) estimates of sub-queries. 
Cardinality here is defined as the number of answers produced by an executed query. 
The optimizer relies on these estimates to determine the order in which the sub-queries are evaluated over the KG to minimize the number of intermediate results, thus, speeding up the query execution.   
One of the challenges in query optimization over KGs is that conjunctive queries typically contain many joins; 
in this context, a join occurs when subgraphs of the query graph are connected by the same variable. 
For example, each pair of edges connected to the same entity can be seen as a join. 
Another challenge is imposed by the semi-structured nature of KGs, which can lead to skewed data distributions and irregular correlations between entities and predicates.  
These are difficult to capture accurately in traditional statistical summaries~\cite{charsets, Stefanoni2018} for different classes of graph queries.

To overcome the limitations of statistical summaries, recent approaches based on Machine Learning (ML) have been proposed to the problem of query cardinality estimation~\cite{Davitkova2022,lss}. 
These models rely on a Neural Network (NN) or Graph Neural Network (GNN) to learn the underlying data distributions in a KG. 
For this, the models are trained in a supervised manner using sampled query graphs and their corresponding cardinalities. 
While it has been shown that these approaches are able to capture arbitrary correlations more effectively than other summaries, they still present several drawbacks. 
First, during learning, these approaches do not leverage individual characteristics of entities and predicates in the KG, i.e., the initial representations are given by vectors that lack meaning or relevant information about the elements. 
Second, some of these approaches do not perform well in \textit{inductive cases}, i.e., when entities occur in queries that were not seen by the model or, in an extreme case, queries that consist of only unseen entities. 
This form of inductive setting occurs when training on query samples that do not cover all entities in the KG.  
Third, the training of the state-of-the-art models requires the optimization of a large number of parameters, which means that a large number of training samples are required to perform accurate predictions.  
This considerably hinders the scalability of state-of-the-art ML approaches to large KGs. 

In this work, we present \textit{GNCE}, \textit{Graph Neural Cardinality Estimation}, a solution to mitigate the drawbacks of the state of the art. 
\textit{GNCE} can be used to predict the cardinality of conjunctive queries over RDF graphs formulated in SPARQL.
\textit{GNCE} is based on a Graph Neural Network~(GNN) as this is an expressive model~\cite{GIN} that can naturally capture the graph structure of the query. 
Furthermore, our approach exploits recent advances in knowledge graph embeddings (KGE)~\cite{Rossi2021} to provide semantically meaningful features of the query elements to the GNN. 
KGE are low-dimensional vectors that represent entities and predicates that capture latent semantic information about the KG. 
KGE provide two key advantages to \textit{GNCE}: to learn data correlations from compact yet meaningful representations, and to generalize to entities unseen in training queries. 
Lastly, as both the meaning and structure of the query graph are largely covered by the KGE and the GNN, \textit{GNCE} can afford to have a relatively simple architecture in terms of the number of learnable parameters. 
This makes \textit{GNCE} applicable to large KGs thus overcoming the third limitation of current solutions. 
In summary, the contributions of this work are:
\vspace{-2mm}
\begin{enumerate}
    \item We present a novel supervised model, termed \textit{GNCE}, for query cardinality estimation over KGs that effectively combines Graph Neural Networks and KG embeddings. 
    \item Our solution is robust in the inductive case, i.e., queries involving unseen entities can be effectively processed by the model. 
    \item We assemble a fast and lightweight ML architecture that is parameter- and data-efficient, i.e., \textit{GNCE} can scale up to large KGs.  
    \item We show that \textit{GNCE} on average outperforms the state of the art using benchmarks from the literature.   
\end{enumerate}

\section{Preliminaries}
\label{sec:preliminaries}

\subsection{Knowledge Graphs}
A knowledge graph (KG) is defined as a directed graph, where nodes and edges are identified with labels~\cite{KG}. 
In addition, nodes can be assigned to classes\footnote{In the literature, sometimes classes are also referred to as \textit{labels} but these should not be confused with the labels/identifiers.}, which correspond to groups of nodes that share some characteristics. 
To represent KGs, there are several data models. 
In this work, we follow the definition of the Resource Description Framework (RDF)~\cite{rdf}, where every statement is modeled as a triple. 
For simplicity, the terms defined in RDF (i.e., IRIs, blank nodes\footnote{The following definitions assume \textit{skolemised} blank nodes, i.e., transformed into IRIs.}, and literals) are treated as labels in the knowledge graph and represented as a universe of terms $\mathcal{U}$. 
%
\begin{definition}[Knowledge Graph]
A knowledge graph $\mathcal{G}$ is a directed, labeled graph, represented as a set of triples, where each triple $t = (s, p, o)$ represents a labeled edge between nodes in the graph. In each triple $t$, $s$ is the subject, $p$ is the predicate, and $o$ is the object, and $s, p, o \in \mathcal{U}$, where $\mathcal{U}$ is a set. The elements of a triple are referred to as atoms.
\end{definition}
Questions over KGs can be formulated as structured queries. 
We focus, in this work, on \textit{conjunctive queries} which can be defined also as a graph where atoms are either instantiated to (bound) \textit{terms}, or \textit{variables} that act as placeholders that can be matched over the KG. 
%
%
\begin{definition}[Query Graph]
\label{def:query}
Consider the set of variables \( \mathcal{V} \) such that \( \mathcal{V} \cap \mathcal{U} = \emptyset \).
A query graph \( Q \) is a directed, labeled graph represented as a set of triple patterns, where each triple pattern \( tp = (s_q, p_q, o_q) \) represents a labeled edge between nodes in the query graph and \( s_q, p_q, o_q \in \mathcal{U} \cup \mathcal{V} \).
\end{definition}

Answers to a query are then found by matching the corresponding query graph over the KG. 
This process is known as \textit{pattern matching}, where atom variables are mapped to elements in the graph. 
A query solution is a subgraph in the KG to which the query graph is homomorphic. 
In particular, in KGs with bound atoms (i.e., no variables in the KG), this is equivalent to simply replacing variables in the query and determining whether the resulting graph is a subgraph in the KG. 
%
%
\begin{definition}[Query Solution]
Given a knowledge graph \( \mathcal{G} \) and a query graph \( Q \), consider a mapping \( \mu: \mathcal{V} \mapsto \mathcal{U} \). 
We denote \( \mu(Q) \) as the resulting set of triples obtained by applying \( \mu(v) \) for each variable \( v \) in the triple patterns of \( Q \). 
If \( \mu(Q) \subseteq \mathcal{G} \), then \( \mu(Q) \) is a subgraph of $\mathcal{G}$ and called query solution to \( Q \) over \( \mathcal{G} \).
\label{def:query_solution}
\end{definition}
The cardinality is the number of solutions to a query:
\begin{definition}[Query Cardinality] The \textit{cardinality} of a conjunctive query $Q$ over $\mathcal{G}$, denoted $||Q||_\mathcal{G}$, is the number of different solutions as of Definition~\ref{def:query_solution}:
\begin{align*}
     ||Q||_{\mathcal{G}} = |\{ \mu: \mu(Q) \subseteq \mathcal{G} \ \land vars(Q)=dom(\mu) \}|
\end{align*}
$vars(Q)$ represents the variables that occur in $Q$, and $dom(\mu)$ denotes the set of variables in $\mu$.
\end{definition}


\subsection{Knowledge Graph Embeddings}
\label{sec:embeddings}

Knowledge graph embeddings (KGE) are low-dimensional vectors of the atoms (entities and predicates) in a KG. 
These vectors encode latent semantic information about the atoms that is not explicitly available in the KG.
For this reason, KGE usually achieve superior results in downstream tasks~\cite{Rossi2021}, e.g., link prediction, entity classification, clustering, etc., in comparison to approaches that only rely on the (explicit) statements and the structure of the KG.   

To compute KGE, existing models for representation learning over KGs comprise an embedding space, a scoring function $f$, and a suitable loss function $l$. The embedding space is typically a vector space over real or complex numbers, with dimensions ranging from dozens to a few hundred. The scoring function $f$ assigns a score to a triple $(s,p,o)$ where a larger score means that the triple is more probable in a KG $\mathcal{G}$ under the model, while lower scores indicate a lower probability.  
The loss function $l$ is used to actually fit the model and drives it to assign high scores to \textit{true triples} -- statements that are assumed correct in the KG -- and low scores to \textit{false triples} -- statements that do not and should not exist in the KG. 
An overview of state-of-the-art KGE is provided by Rossi et al.~\cite{Rossi2021}.

Due to differences in their components and learning process, different KGE models are able to capture different latent aspects of KGs, which makes certain KGE more suitable for specific tasks. 
For instance, some KGE models (e.g., translational models~\cite{Rossi2021}) learn from individual triples in the KG, which are well suited for link prediction. 
Other models (e.g., RDF2Vec~\cite{ristoski2016rdf2vec} or Ridle~\cite{DBLP:conf/cikm/WellerA21}) can learn from larger structures like subgraphs, which are suited for tasks that account for the semantic similarity and relatedness of atoms like entity type prediction or clustering.  
In this work, we are interested in the latter group of KGE models, as learning from subgraphs allows for capturing the correlations of atoms in the KG effectively. 
This is achieved with RDF2Vec~\cite{ristoski2016rdf2vec} which implements the \textit{skip-gram model}~\cite{word2vec} based on random walks of atoms in a KG. 
A random walk over a KG is generated by traversing the graph starting from an arbitrary node, following an edge to another node, and so on. 
Given a random walk of atoms $w_1,...,w_W$ (which includes nodes and predicates) in a KG of length $W$, the objective of RDF2Vec is to maximize the following log probability (i.e. minimize its negative, which can be seen as the loss function in KGE):
\begin{equation}\label{rdf2vec1}
    \frac{1}{W} \sum_{t=1}^W \sum_{-c \leq j \leq c, j \neq 0} \log p (w_{t+j}|w_t)
\end{equation}

That is, given a target atom $w_t$, RDF2Vec tries to maximize the probability of all atoms (summation over $j$) that are at most $c$ hops away in the walk. 
RDF2Vec does that for every atom in the random walk (summation over $t$). 
The conditional probability of an atom $w_o$ given another atom $w_i$, which can be seen as the analog to the scoring function in other KGE methods, is given as:
\begin{equation}
    p(w_o|w_i) = \frac{\exp (v'^\top_{w_o} v_{w_i})}{\sum_{w=1}^{|V \cup E|} \exp (v'^\top_{w} v_{w_i})}
\end{equation}

Here, $v_{w_j}$ denotes the input vector (embedding) of atom $w_j$, while $v'_{w_j}$ denotes the so called output vector of $w_j$. Those have the same dimension as the input vector but are not used as the final embeddings. They are introduced to make the model more expressive and have different representations for target atoms and context atoms.
As we can see, the unnormalized conditional probability is estimated as the exponent of the dot product between the input and output vectors of both atoms. 
This dot product gets maximized if the angle between the two vectors is zero. 
That is the case if they are the same vector or lay on the same 1-D subspace. 
Thus, the optimization of Eq.~(\ref{rdf2vec1}) drives embeddings of atoms with the same context to be close to each other. That is especially true for atoms that occur together in the graph, i.e. when the joint probability of observing them is high.
Note that a normalized probability is obtained by dividing by the sum of all possible probabilities (often termed partition function). That sum is difficult to calculate and in practice approximated~\cite{ristoski2016rdf2vec}.

The important takeaway point here is that the embeddings generated by RDF2Vec encode the atom's relatedness and correlate with the joint probability of observing them. At first glance, it might appear as if the embeddings encode the sampled paths, but since a single entity participates in many walks, rich contextual information from the neighboring nodes and edges is encoded. Our hypothesis is thus that RDF2Vec embeddings are suitable for atom representation in estimating query cardinalities over KGs.


\subsection{Graph Neural Networks}
\label{sec:gnn}

A Graph Neural Network (GNN) is a special form of Neural Network (NN) that is tailored to graph-structured data.  
The distinct feature of GNNs is that they are invariant with respect to a permutation of the nodes of the input graph~\cite{Bronstein2021}. 
This means that the output of the model is the same even if the atoms in the graph are reordered. 
Permutation invariance makes GNNs parameter and data efficient~\cite{Bronstein2021}. 
The model is parameter efficient as it needs fewer learnable parameters because it is not necessary to learn different permutations of the same graph structure. 
It is also data efficient as the model needs a relatively low number of examples to perform well. 
One typical task GNNs are used for is graph-level prediction. Here, given a graph as input, the GNN predicts quantities that hold for the whole graph. E.g. if a molecular graph is toxic or a social network is racist. Important for us, predicting the cardinality given a query graph can be considered a graph-level prediction.

GNNs work under the message-passing framework, where in every layer $k$ in the GNN the input representations of the nodes are updated based on \textit{messages} from all connected neighbors. 
The most generic \textit{message-passing function} for a node is given by~\cite{Pyg1}:

\begin{equation}
    x_i^{(k)} = \gamma^{(k)}\left(x_i^{(k-1)}, \square_{j \in \mathcal{N}(i)} \phi^{(k)}\left(x_i^{(k-1)}, x_j^{(k-1)}, e_{j,i}\right)\right)
\end{equation}

Here, $\gamma$ and $\phi$ are differentiable functions, and $\square$ is a differentiable as well as permutation invariant function. $\mathcal{N}_i$ is the set of direct neighboring nodes of the $i$-th node.
For a node $x_i$, its previous representation denoted $x_i^{(k-1)}$ gets non-linearly transformed to a new representation $x_i^{(k)}$ by combining $x_i^{(k-1)}$ with neighboring nodes' representations as well as the edges between them (the messages). 
In order to perform graph classification, the individual node features of all nodes need to be combined into a fixed-length vector. 
For that, a function that is permutation-invariant and works with a varying number of nodes in the graph is suitable. 
Practically used examples here are max-pooling, mean-pooling, or sum-pooling. The resulting vector represents the whole graph and can be used to predict the desired graph-level quantities~\cite{Wu2019}. 

The initial representation of a node, $x_i^0$, depends on the use case. 
Possible options are one-hot or binary encoding to denote the \textsc{id} of the node or more informative features. E.g. if the node represents a person in a social graph, one dimension could represent the height of the person, another the age, etc. More meaningful features are superior as they provide crucial information to the GNN to solve the desired problem. 
They also free the GNN from learning internal representations of the nodes, as would be the case for one-hot or binary encoding (as the feature does not carry any meaning except for the identity of the node) where any semantics of the node would need to be learned from the data and integrated into the model. 
As KGE provide semantic information about the given entity, they are a promising option as initial node features.
\section{Related Work}\label{sec:related_work}
Cardinality estimation is a longstanding research problem in relational databases    
and proposed solutions have been taken as starting points for graph-structured data like KGs. 
Traditional approaches rely on summaries of the dataset but, more recently, learning-based solutions have been proposed.    
In the following, we discuss relevant work in both 
relational and graph-structured data. 

\paragraph{Approaches for Relational Databases}
Traditionally, cardinality estimation is based on sampling and histograms. 
Histograms can theoretically provide the full joint probability over attributes and join predicates, however, they are computationally prohibitive, since they grow exponentially with the number of attributes. One therefore often employs independence and uniformity assumptions over the data, which leads to very high cardinality estimation errors. 
Those can be reduced by computing an upper bound on the cardinality estimate~\cite{upper_bound_relational}.
Other approaches to overcome this, e.g., sampling approaches~\cite{Cormode2011,Vengerov2150,Li2016}, have been devised that do not make any independence assumptions about the data. 
By evaluating the query over a sample from the data, the full joint probability is evaluated on that sample.  
As the sample size approaches the size of the real data, the true cardinality over the data is approached. 
Yet, in practice, the sample size is usually much smaller than the dataset to compute the estimates in a reasonable time. 
Therefore, the accuracy of the estimate highly depends on the sampling technique as well as the data distribution, and the sampling might need to run long in order to generate a good estimate. 
Approaches in the context of sampling are, e.g., \textit{Correlated Sampling}~\cite{Vengerov2150}, \textit{Wanderjoin}~\cite{Li2016}, and \textit{Join Sampling with Upper Bounds (jsub)}~\cite{jsub}. 
\textit{Correlated Sampling}~\cite{Vengerov2150} improves over simple Bernoulli sampling by sampling according to hashed values of the attributes, thus,  reducing the variance of the estimate. 
\textit{Wanderjoin}~\cite{Li2016} represents the query and database as a query graph and data graph and estimates cardinalities by performing random walks on the data graph in coherence with the query graph. 
\textit{Join Sampling with Upper Bounds} performs sampling similar to \textit{Wanderjoin} but extends the estimate to provide an upper bound of the cardinality. The methods are summarized in~\cite{Park2020}.
For a comprehensive introduction to sampling and histogram approaches for cardinality estimation, consider the survey by Cormode et al.~\cite{Cormode2011}.

To circumvent the exponential explosion of compute- and memory requirements by statistical summaries, learning approaches have been proposed.  
Sun et al.~\cite{Sun2021} group these approaches into the so-called \textit{data models} and \textit{query models}. 

Data models first aim at predicting the probability of the data and then estimate the query cardinality by sampling from the learned probability distribution.  
Some examples of these solutions are based on Bayesian Networks~\cite{Getoor2001,Tzoumas2011,Halford2019}. 
All these approaches aim at representing the joint probability by factorizing it into a product of smaller conditional probabilities~\cite{Halford2019}. 
The problem is that, in practice, the factorization can require capturing joint correlations of many attributes in the data, thus, growing the network size also exponentially. 
For this reason, in practice, Bayesian Networks still make independence assumptions to a great extent~\cite{Halford2019}.
To overcome these limitations, \textit{Sum Product Networks} implemented in DeepDB~\cite{Hilprecht2020} are used as graphical models to approximate the data joint probability. 
Compared to Bayesian Networks, the memory of Sum Product Networks grows polynomial w.r.t. the database size.

Query models, instead, aim at predicting the cardinality given a query. 
Most of these approaches are based on Neural Networks (NNs). 
For example, Kipf et al.~\cite{kipf2018learned} present \textit{Multi-Set Convolutional Networks} (MSCN), which represents queries by using table, join, and predicate sets. 
Woltmann et al.~\cite{Woltmann2019} also propose an NN-based approach but this focuses on local parts of the data.  
The solution by Zhao et al.~\cite{zhao2022lightweight} is based on NN Gaussian Processes (NNGP) to handle uncertainty to improve the accuracy of predictions. 
While the latter solutions are closer to our work, these NNs are tailored to queries over relations and cannot directly be applied to graphs.

\paragraph{Approaches for Graph-Structured Data}
Early works on cardinality estimation for graphs take inspiration from solutions for relational databases by using histograms and assuming attribute independence~\cite{Stocker2008,shironoshita,Neumann2009}. 
More recent approaches try to model correlations between structures of the query. 
Neumann et al.~\cite{charsets} propose \textit{Characteristic Sets} (\textit{CSET}), which are synopses that count the number of entities with the same set of predicates. 
\textit{CSET} provides exact cardinality estimates for star-shaped \texttt{DISTINCT} queries where all predicates are instantiated and objects are unbounded (variables). 
For other classes of queries, Neumann et al.~\cite{charsets} decompose the query into star-shaped subqueries, calculate their cardinality using \textit{CSET}, and estimate the final cardinality by assuming independence between the sub-queries. 
\textit{SUMRDF}~\cite{Stefanoni2018} is also based on graph summarization and computes \textit{typed} summaries where nodes that belong to the same class\footnote{In RDF KGs, the class of a node is provided with the predicate \texttt{rdf:type}.} and have similar predicate distributions are grouped together.  
In addition to summaries, sampling-based approaches have also been proposed for KGs. 
For example, \textit{impr}~\cite{Chen2017} estimates the query cardinality through random walks on the data that match the query graph.   
Lastly, approaches based on Bayesian Networks have also been proposed for KGs~\cite{Huang2011}.
All the solutions aforementioned are tailored to RDF KGs. 
However, similarly to their relational counterparts, the structures computed with these approaches can quickly grow w.r.t. the dataset size or are only accurate for specific classes of queries. 

\begin{table*}[t!]
  \caption{Characteristics of the learned approaches}
  \label{tab:sota}
  \footnotesize
  \resizebox{0.99 \textwidth}{!}{%
  \begin{tabular}{llllllll}
    \toprule
    \textbf{Approach} & \textbf{ML Model} & \textbf{Atom Representation}& \textbf{Model Size} & \textbf{Embeddings} &  \textbf{Perm. Inv.} &  \textbf{Assumptions} & \textbf{Inductivity (Empirical)}\\
    \midrule
    \textit{LMKG}~\cite{Davitkova2022}  &   NN & Binary Encoding & $O(|V|+|E|)$ & - &  \xmark & -- & Entities, Query Shapes\\
    \textit{LSS}~\cite{lss}  &   GNN & ProNE~\cite{prone} & Constant & $|V|$ & \cmark &  Typed entities, undirected graph & Entities, Query Shapes\\
    \midrule
    \textit{GNCE} (Ours)  &  GNN & RDF2Vec~\cite{ristoski2016rdf2vec} &Constant & $|V|+|E|$& \cmark & --
    & Entities, Query Shapes\\
    \bottomrule
  \end{tabular}
  }
\end{table*}

Most similar to our work are \textit{LMKG}~\cite{Davitkova2022} and \textit{LSS}~\cite{lss}, which are approaches based on Machine Learning (ML). 
Table~\ref{tab:sota} presents an overview of these solutions, which are discussed in detail next. 

\textit{LMKG}~\cite{Davitkova2022} is a supervised model\footnote{There is another version of \textit{LMKG} with unsupervised learning, however, its performance was inferior in comparison to its supervised counterpart. Therefore, we focus only on the supervised setting of \textit{LMKG}.} for learning cardinalities using Neural Networks (NNs) consisting of several nonlinear fully connected layers. 
The representation of the KG and queries in \textit{LMKG} works as follows. 
First, each atom in the KG is assigned an \textsc{id} and transformed into a binary vector representation of that \textsc{id}.  
Then, a graph query with $n$ nodes and $e$ edges is represented with an adjacency tensor as well as featurization matrices to relate atoms in the query to atoms in the KG.  
In practice, \textit{LMKG} assumes that the adjacency tensors and featurization matrices of all queries have the same dimensions, meaning that $n$ and $e$ are set to the size of the \textit{largest} query in a workload; this makes the resulting model possibly large.
A possible solution is to train several models for different query sizes~\cite{Davitkova2022}, yet, this induces a managing overhead. 
After the representations are computed, the input to the NN is the concatenation of the flattened query representations. 
Because of this, \textit{LMKG} is not permutation invariant, as the same query where triple patterns are provided in a different order yields different representations. 
That means that \textit{LMKG} needs a higher number of parameters and training data points to capture these permutations and produce accurate predictions.  

\textit{LSS}~\cite{lss} applies a Graph Neural Network (GNN) to the graph query. 
For each node in a query, \textit{LSS} computes the tree induced by a $l$-hop BFS (Breadth-First Search) starting at that node. 
Then, the GNN is applied to all those trees; the results are aggregated by an attention mechanism to finally predict the cardinality. 
While \textit{LSS} has been empirically proven to be effective for cardinality estimations over (regular) graphs, the initial representation of query nodes in \textit{LSS} relies on two aspects that make it less suitable for KGs.  
First, \textit{LSS} computes the initial representation of the nodes in the graph query with ProNE~\cite{prone} embeddings, which are tailored to graphs and not KGs, as they do not consider typed and directed edges.  
Second, each atom is represented with the embedding that results from summing up the embeddings of their corresponding classes in the KG. 
A problem with the summed class representation is that queries with the same shape but different entities that belong to the same classes have the same representation. 
This is not ideal for cardinality estimation as the selectivity of these queries can greatly vary due to the different entities mentioned in the query. 
Additionally, \textit{LSS} employs a message-passing function that treats the graph as undirected. 
Furthermore, the attention mechanism introduces a large number of parameters (due to the $\mathbf{Q, K, V}$ matrices and MLP of an attention head~\cite{vaswani2017attention}), making the resulting model large (with over 2 million parameters in our experiments). 
\textit{LSS}  also supports active learning, which primarily offers accelerated training. By focusing on poorly performing queries, the model can improve its performance without the need for exhaustive training over the entire dataset.
Yet, active learning is orthogonal to the architecture and could be applied to our model identically.

\textit{NeurSC}~\cite{neursc} applies several GNNs to both the query graph as well as subgraphs of the data graph. 
The latter provides additional information, but it is more costly since the graph needs to be sampled for every query. 
Furthermore, \textit{NeurSC} is tailored to undirected graphs without edge labels, which is not suitable for KGs. 
For this reason, we consider that the closest works to ours based on ML are \textit{LMKG} and \textit{LSS}. 
Furthermore, \textit{LMKG} and \textit{LSS}, as well as our method \textit{GNCE}, do not make any independence or uniformity assumptions about the data, as the underlying NN can in principle approximate any skewed distribution.

Lastly, in query processing over KGs, there are related research problems that have been investigated, e.g., worst-case join order optimization~\cite{HoganRRS19} or the prediction of query execution performance~\cite{ZhangSQTY18,HasanG14,casalssparql}. 
However, these problems greatly differ in nature and proposed solutions to the ones tackled in this paper, therefore, they are considered out of the scope of this work.   
\section{Our Approach}
\label{sec:approach}
We model the task of cardinality estimation over a Knowledge Graph (KG) as a regression process, where we directly use a predictive model to estimate the cardinality of a given query graph. Concretely, we focus on conjunctive queries formulated in SPARQL over RDF KGs.

Our model \( f \) aims to predict the cardinality of a query \( Q \) over the Knowledge Graph \( \mathcal{G} \). This prediction is based on the query's feature representation, denoted as \( \mathcal{Q} \) (the specifics of which are detailed later in the text):
\begin{equation}\label{eq:cardinality_probability}
    f(\mathcal{Q}) = \widehat{||Q||}_\mathcal{G} \approx ||Q||_\mathcal{G}
\end{equation}
Here, \( f(\mathcal{Q}) \) yields the estimated cardinality \( \widehat{||Q||}_\mathcal{G} \), which approximates the true cardinality \( ||Q||_\mathcal{G} \) of the query within the graph.
\begin{figure}[t!]
  \centering
  \includegraphics[width=0.6\textwidth]{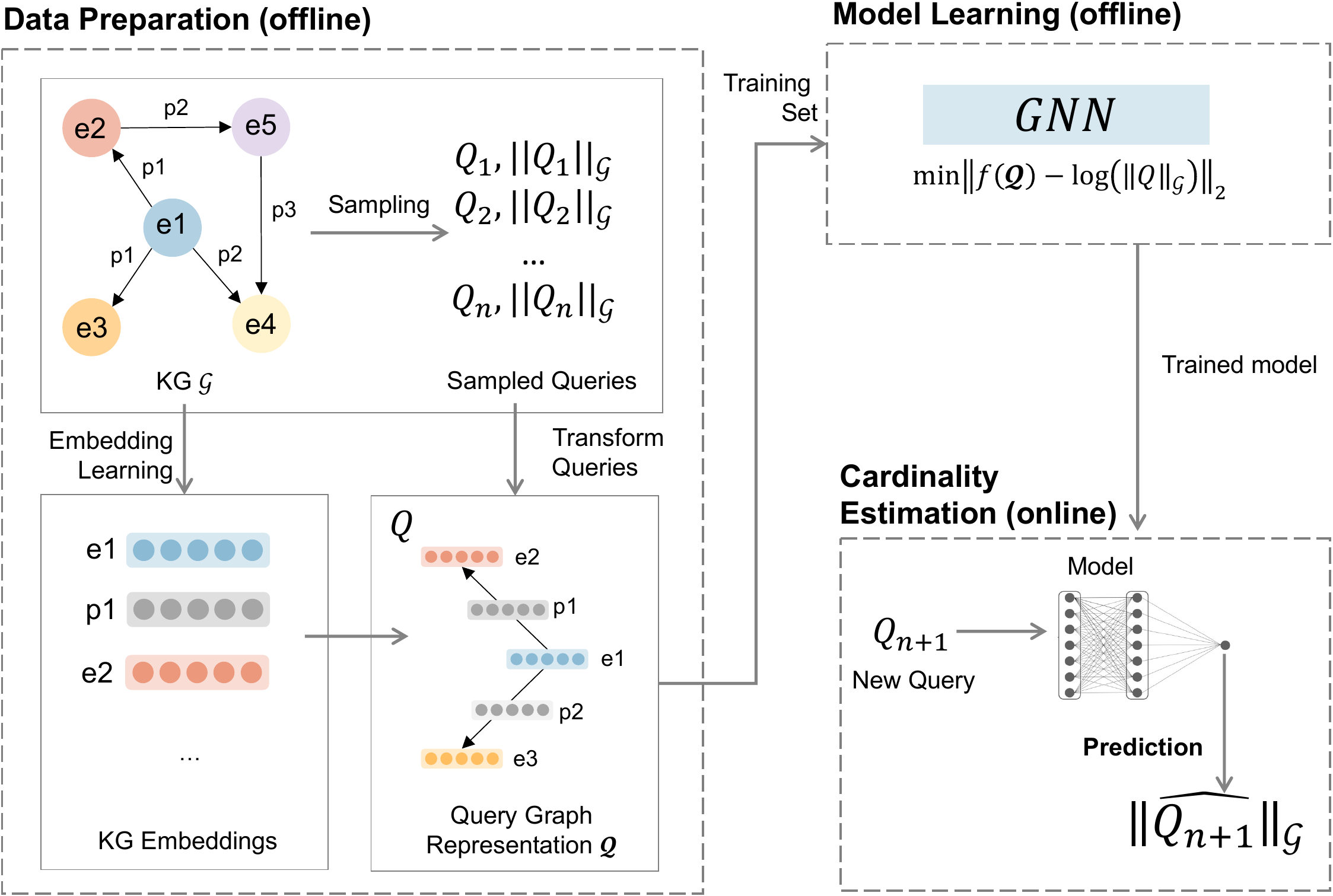}
  \caption{GNCE Overview. For a KG, embeddings are learned for existing atoms. Next, a GNN is trained using sampled queries, which are represented with the embeddings. Finally, the trained model estimates the cardinality of new queries.}
  \label{fig:framework}
\end{figure}


To approximate the true cardinality as outlined in Eq.~(\ref{eq:cardinality_probability}), we choose a neural network representation, due to its capacity for learning complex functions~\cite{maiorov1999lower}. Considering that queries over Knowledge Graphs (KGs) inherently form graph structures (cf.  Definition~\ref{def:query}), employing a Graph Neural Network (GNN) is a fitting choice. This selection ensures permutation invariance within the query graph and also enhances parameter and data efficiency.

Figure~\ref{fig:framework} provides an overview of our approach. In the initial phase, \textit{GNCE} prepares the data necessary for training the machine learning model. This preparation phase, detailed in Section~\ref{subsub:featurization}, involves sampling queries from $\mathcal{G}$, computing Knowledge Graph Embeddings, and building meaningful query representations. Subsequently, \textit{GNCE} employs a GNN to learn a model from these query representations. The learning objective is to minimize the discrepancy between the model's predicted cardinality and the actual cardinality of the queries, as discussed in Section~\ref{sec:model_training}. Upon successful training, the GNN model is capable of predicting the cardinality for new queries unseen during the training phase (cf. Section~\ref{sec:prediction}).

\subsection{Data Preparation}
\label{subsub:featurization}
The goal of the data preparation is to obtain meaningful representations that can be used to learn a model for cardinality estimation. 
The data preparation in \textit{GNCE} is carried out offline and consists of (1) obtaining queries from the KG to use for training, 
(2) computing embeddings for atoms that occur in the queries, 
(3) building the featurization of each query using (1) and (2).
This process is described in the following.

\textit{GNCE} first obtains a training set of queries with their true cardinality from a given KG $\mathcal{G}$. 
For this, \textit{GNCE} randomly samples queries from $\mathcal{G}$ with different structures, e.g., different shapes like stars or paths, a varying number of triple patterns, and different bound terms. 
During this process, it is important to sample uniformly across the KG to cover the distribution of data as best as possible and not only fit some regions of likely user queries.
Then, for each sampled query $Q$,  
the individual occurrences $occ(x)$ in $\mathcal{G}$ of each bound atom $x$ appearing in $Q$ is calculated. 
That is, $occ(x)$ is the number of triples in $\mathcal{G}$ that atom $x$ occurs in any position, i.e., subject, predicate, or object. 
Those quantities will be informative to the approximation of the overall query cardinality: the higher the number of atom occurrences, the higher the probability that a subgraph of $\mathcal{G}$ will match the query graph $Q$. 
As the $occ(x)$ values can be easily obtained using simple queries, it is much better to provide them to the model instead of the model having to learn them. 

Next, \textit{GNCE} calculates embeddings $e(x)$ for atoms (nodes and predicates) in $\mathcal{G}$ using RDF2Vec~\cite{rdf2vec}.  
Since, as explained in Section~\ref{sec:embeddings}, RDF2Vec embeddings encode relatedness and joint probability, we choose it as our embedding method.
With RDF2Vec, nodes (including IRIs and literals) and predicates are treated as elements in the random walks, therefore, RDF2Vec is able to produce embeddings for all types of atoms in the KG.
In this way, \textit{GNCE} also supports literals, even though they are treated equally to IRIs (similar to all compared methods). 
That is, the ordinal similarity between literals in the different data types is not considered, and we leave a special treatment (e.g. \cite{DBLP:conf/semweb/KristiadiKL0F19, preisner2023universal}) for future work.

Moreover, \textit{GNCE}
only trains the RDF2Vec embeddings on the entities occurring in the sampled queries instead of the full KG, which greatly improves training speed for large graphs (see \cite{rdf2veclight} for details). This is justified as we observed that the resulting model performance is similar when training on the whole graph. However, in a production setting, one might want to calculate embeddings for all entities, or at least for those frequently occurring in queries. Note that the process of embedding learning happens prior to training the model for cardinality estimation, and the embeddings of existing entities are fixed after this. In Section~\ref{sec:prediction}, we discuss how to handle new entities after training. We assume that all predicates that will appear during testing and later estimation on new queries are covered in the training set.

Lastly, for each sampled query $Q$, \textit{GNCE} computes a representation $\mathcal{Q}$.  
The representation $\mathcal{Q}$ consists of replacing each atom in the query graph $Q$ with a final input feature that is meaningful for learning the cardinality model.
Here, \textit{GNCE} builds the final input feature $i(x)$ for an atom $x$, which can be a bound or a variable atom. 
For a bound atom $x$, \textit{GNCE} concatenates its corresponding embedding $e(x)$ and the calculated occurrence $occ(x)$, i.e., $i(x) = e(x) || occ(x)$. 
The case of variable atoms in the query graph requires a special vector representation. 
For each query individually, \textit{GNCE} randomly assign an increasing numerical \textsc{id} to each variable (e.g. \textit{?s1} gets \textsc{id} 2, \textit{?p1} gets \textsc{id} 1, etc.). 
Then, \textit{GNCE} builds the vectors $i(x)$ for a variable atom as follows. 
The first dimension of the vector is set to the \textsc{id}. 
The rest of the dimensions are all set to $1$. By that, the model is informed about which atoms are variables, and which variables are the same. 
Since the assignment is random, the model does not falsely learn that a certain \textsc{id} corresponds with a certain position or label.
This is different from other approaches~\cite{Davitkova2022,lss}, where distinct variables in a query are encoded with 0s and treated as if they were the same variable.
Finally, the featurization $\mathcal{Q}$ is then given by $i(x), \forall x$. In that way, \textit{GNCE} jointly encodes local neighborhood and similarity to other atoms, as well as statistical information about the entity.

\subsection{Model Learning}
\label{def:model}


In this section, we present the \textit{model definition} that explains how \textit{GNCE} learns from the representations $\mathcal{Q}$ obtained during data preparation, followed by the details of the \textit{model training} that describes how \textit{GNCE} implements supervised learning using true query cardinalities to perform predictions.    
\begin{figure*}[t!]
  \centering
  \resizebox{0.99\linewidth}{!}{\includegraphics{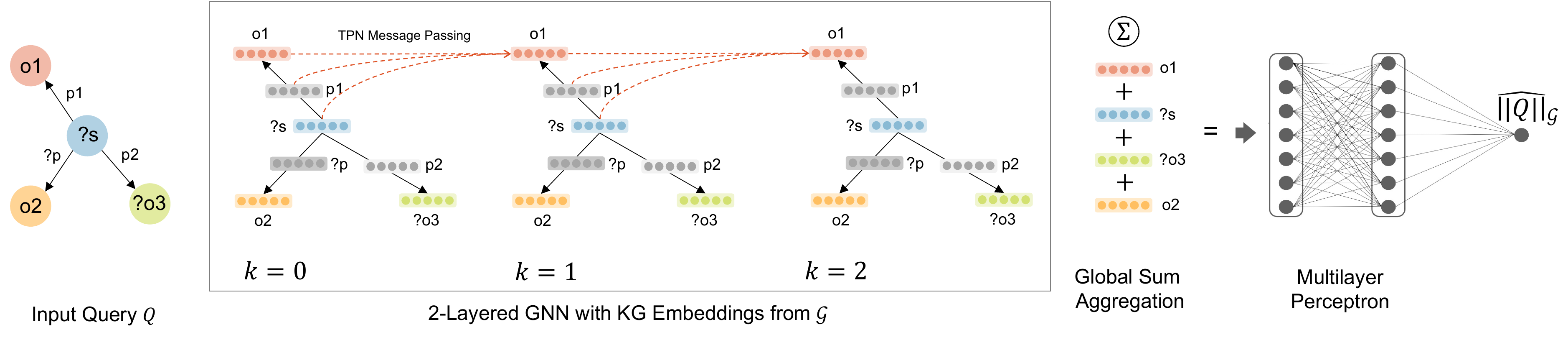}}
  \caption{Architecture of the GNN for cardinality estimation. An initial query is represented by entity- and predicate embeddings and the connectivity between those atoms. For clarity, the TPN aggregation is only shown for one node. After node aggregation, the complete query graph is represented by summing all node embeddings. Finally, the query cardinality is estimated by mapping the query representation through a multilayer perceptron to a single dimension.}
  \label{fig:GIN}
\end{figure*}

\paragraph{Model Definition.} 
\label{sec:model_definition}
Figure~\ref{fig:GIN} shows the  \textit{GNCE} architecture that consists of a Graph Neural Network (GNN) that implements Message Passing Layers with a Multi-layer Perceptron (MLP).

The representation $\mathcal{Q}$ is the input to our GNN ($k=0$). 
First, two Message Passing Layers ($k=1, k=2$) are applied to the input. 
In a preliminary study, we found that using more layers does not improve the model's performance. 
Possible explanations for this are that (i) the query graphs are rather small so two steps of message-passing are sufficient, and (ii) fewer layers avoid the oversmoothing problem~\cite{oversmoothing}, where node representations become increasingly similar, which deteriorates the model performance.

For the Message Passing Layers in the GNN, we propose a custom  Message Passing function. 
Our function, which we term \textit{TPN Message Passing}, is a modulation of the \textit{GINECONV}~\cite{Hu2019} message passing function defined as
\begin{equation}\label{eq:GINECONV}
    x_i^{(k)} = h_\theta^{(k)}\left( (1+\epsilon) x_i^{(k-1)} + \sum_{j \in \mathcal{N}(i)}
    \mathrm{ReLU}(x_j^{(k-1)} + e_{j,i})\right)
\end{equation}

We initially evaluated several different Message Passing Layers but found \textit{GINECONV} to be superior to the others. 
This is in line with the theoretical findings by Xu et al.~\cite{GIN}, which states that \textit{GINECONV} is strictly more expressive than other Message Passing Layers.
It also aligns with results from \textit{LSS}~\cite{lss}, which also uses \textit{GINECONV} as the Message Passing Function.

In Eq.~(\ref{eq:GINECONV}), each $h_\theta^{(k)}$ is a multilayer perceptron (MLP) with a ReLU Activation function. 
$\mathcal{N}(i)$ are all neighboring nodes of node $i$ with corresponding edges $e_{j,i}$. 
The parameter $\epsilon$ is an optionally learnable parameter that defaults to zero. 
Therefore, this function first sums up the neighboring nodes and edges individually, then aggregates those and the current representation of the node $x_i^{(k-1)}$ again through summation to finally transform them nonlinearly through $h_\theta$. Unfortunately, this Message Passing function treats the graph as undirected, as it aggregates the neighbors invariant under the direction of the edge. 
However, in cardinality estimation, the edge direction can be important information: 
The triples $(s \rightarrow p \rightarrow o)$ and $(o \rightarrow p \rightarrow s)$ can have vastly different cardinalities. 
One option to make the message passing directional is to only aggregate incoming edges at each node, i.e., instead of summing over $\mathcal{N}(i)$, the summation happens over $\mathcal{N}^{+}(i)$. However, we found that this yields worse results than just summing over $\mathcal{N}(i)$, presumably because it restricts information flow. Hence, we use the following customized message-passing function:

\begin{definition}[TPN Message Passing]
    The Triple Pattern Network (TPN) Message Passing transforms a node representation $x_i^{(k-1)}$ by separately transforming incoming messages (from $\mathcal{N}^+(i)$) and outgoing messages (from $\mathcal{N}^-(i)$) like
    
    \begin{align}
     x_i^{(k)} = h_\theta^{(k)}  \biggl( x_i^{(k-1)} \ & + \sum_{j \in \mathcal{N}^+(i)}
    \mathrm{ReLU}(\mathbf{W}(x_i^{(k-1)}||e^{j,i}||x_j^{(k-1)})) \ + \nonumber\\
    &\vspace{-3em} + \sum_{j \in \mathcal{N}^-(i)}
    \mathrm{ReLU}(\mathbf{W}(x_j^{(k-1)}||e^{i,j}||x_i^{(k-1)})) \biggr)  \nonumber
\end{align}
\end{definition}

This function makes a case distinction between incoming and outgoing edges. 
Instead of aggregating the node and edge features through summation, the function concatenates them ($||$) and linearly projects them to the same dimensionality as $x_i$ using the matrix $\mathbf{W}$. 
Note that it is not necessary to have two different matrices $\mathbf{W}^+$, $\mathbf{W}^-$, since $\mathbf{W}$ always receives the same encoding of a triple pattern $\mathbf{O||P||S}$, where $\mathbf{O}$ is the feature of the node where the edge points to, $\mathbf{P}$ is the edge feature between the nodes, and $\mathbf{S}$ is the feature of the node where the edge originates. 
Lastly, the MLP $h_{\Theta}^{(k)}$ receives as input sums of triple pattern representations.
After the graph nodes have been transformed by the two Message Passing Layers, they need to be aggregated into a fixed-length vector. 
For that, \textit{GNCE} applies a \textit{Global Sum Aggregation} to them. That is, the representations of the nodes are summed up dimension-wise, resulting in a single vector $x$ with the same dimensionality as the node embeddings: $x = \sum_l x_l^{(k=2)}$.

The last step of the \textit{GNCE} model is to transform this vector $x$ through a final 2-layer MLP (with a ReLU activation in the hidden layer and linear activation in the final layer) into one single dimension (i.e., one value). 
This value represents the predicted cardinality of the query by \textit{GNCE}.

\paragraph{Model Training.}
\label{sec:model_training}
To learn the cardinality of queries, \textit{GNCE} follows a supervised approach, using the query representations $\mathcal{Q}$ (processed with the GNN as previously described) and the true cardinalities $||Q||_\mathcal{G}$. 
The goal during training is to minimize the difference between the predicted cardinality, denoted $f(\mathcal{Q})$,  and the true value. 
For that, as a loss function, \textit{GNCE} employs the Mean Squared Error (MSE) between our model output $f(\mathcal{Q})$ and the logarithm of the true cardinality $||Q||_\mathcal{G}$: $\mathcal{L} = ||f(\mathcal{Q}) - \log(||Q||_\mathcal{G})||_2$. 
Hence, the model learns to predict logarithms of cardinalities, which stabilizes the training due to the large range of possible cardinalities. 
Although the model is evaluated in terms of q-Error, we found that using MSE as a loss function yields better results than using q-Error directly. 
We train the model with a batch size of $32$ and for 50 epochs in all experiments. For optimization of the model, we use the Adam optimizer~\cite{Kingma2014} with a learning rate of $e^{-4}$. 
The model training is carried out offline\footnote{Related approaches also pre-compute statistics or samples offline.}, and the learned model can then be used to perform online predictions on new queries that were not part of the training sample, as described next.  

\subsection{Cardinality Estimation on New Queries}
\label{sec:prediction}
The learned model can now be used to predict the cardinality of new queries. 
This prediction is computed online, as a query $Q_{n+1}$ is provided to the approach.  
For this, \textit{GNCE} identifies the bound atoms mentioned in $Q_{n+1}$, and retrieves the corresponding embeddings which have been pre-computed. 
For variable atoms, \textit{GNCE} builds their corresponding vectors, to obtain the representation $\mathcal{Q}_{n+1}$. 
This representation is provided to the trained model, to compute the prediction $||\widehat{Q_{n+1}}||_\mathcal{G}$.  
\paragraph{Model Learning and Deployment in Production Environments.}
We now describe how \textit{GNCE} can be implemented in a real-world environment. 
The KG $\mathcal{G}$ is assumed to be stored in a triple store.
To obtain sample queries, \textit{GNCE} generates random queries that are executed over the triple store, either directly or via a SPARQL endpoint.
Metadata of the queries such as the true cardinalities and the $occ$ values are also obtained by \textit{GNCE} querying the triple store. 
Then, for the atoms that occur in the sampled queries, \textit{GNCE} generates random walks –also querying the triple store– which are used to learn the RDF2Vec embeddings.
The pre-computed embeddings are stored in a fast-accessible storage system, e.g., a vector database. 
Then, \textit{GNCE} trains the GNN model which is compiled and deployed for production. 
When a new query arrives, the relevant embeddings are retrieved from the vector database to build the query's featurization, and the GNN directly predicts the cardinality.


An important aspect within production environments is the addition of new entities to the KG. 
If a new entity in a query does not yet have a corresponding embedding, \textit{GNCE} generates a random embedding to represent the new entity. Theoretically, this can deteriorate the performance as the model is making prediction on an unknown entity. Still, in practice, the \textit{GNCE} performance remains satisfactory (see Section~\ref{sec:inductive_case}). This ensures that the approach will run for any query, regardless if all embeddings already exist.
Eventually, after a certain period or after a certain number of new entities are added, \textit{GNCE} can learn the embeddings for new entities offline. 
In this process, \textit{GNCE} also updates the occurrences $occ(x)$ for atoms related to the new entities; yet, this is a lightweight operation. Afterward, \textit{GNCE} adds the new embeddings to the vector database, making them available to the deployed GNN model.
Please note that all approaches that rely on pre-computed data (e.g., summaries, samples, or ML models) are also affected by changes in the KG, and they also need to update their structures to cover new entities.
\section{Experiments}
\label{sec:experiments}

We empirically investigate 
the effectiveness of the studied approaches (\S\ref{sec:star}, \S\ref{sec:path}, \S\ref{sec:complex}),  
the generalization of the approaches to unseen entities, i.e.,  the inductive case (\S\ref{sec:inductive_case}),  
the runtime efficiency of the approaches (\S\ref{sec:efficiency}), 
and ablation study of GNCE (\S\ref{subsec:ablation}), 
and an application of GNCE in a query optimizer (\S\ref{sec:optimization}).
The raw datasets, queries, results, and code are available on GitHub\footnote{\url{https://github.com/DE-TUM/GNCE}}.

\subsection{Experimental Setup}
\label{sec:settings}

\subsubsection{Datasets and Queries.}
For datasets, we use SWDF~\cite{swdf}, LUBM \cite{lubm}, YAGO~\cite{yago}, and a subset of Wikidata~\cite{wikidata5m} that consists of 20 million triples.
\begin{table}[h!]
\caption{Dataset characteristics}
\label{tab:datasets}
\centering
\footnotesize
\begin{tabular}{|l|r|r|r|r|r|}
\hline
KG & Triples & Entities & Predicates & Classes & Typed Entities \\
\hline
SWDF & 242256 & 76711 & 170 & 118 & 28\%\\
\hline
LUBM & 2688849 & 664048 & 18 & 15 & 66\%\\
\hline
YAGO & ~58M & ~13M & 92 & ~189K & 98\%\\
\hline
Wikidata & ~21M & ~5M & 872 & ~19K & 84\%\\
\hline
\end{tabular}
\end{table}

For LUBM and SWDF, we use the same datasets as Davitkova et al.~\cite{Davitkova2022}, for easier comparison to previous results from the literature. 
For YAGO, we use the dataset provided by the benchmark tool \textit{G-CARE}~\cite{Park2020}. 
LUBM is a synthetic benchmark modeling a university domain. 
SWDF is a small size but dense real-world dataset that models authors, papers, and conferences. 
Lastly, YAGO and Wikidata are large cross-domain KGs.  
Table~\ref{tab:datasets} summarizes the statistics about these datasets.  
For the queries, we generated\footnote{\url{https://github.com/DE-TUM/rdf-subgraph-sampler}} star and path queries as reported in Table~\ref{query_counts}. 
We additionally generated more complex shapes over YAGO and retrieved real queries for Wikidata (detailed in Section \ref{sec:complex}). 
The cardinality of the queries is in the range $[e^0, e^6]$ for LUBM and SWDF, and $[e^0, e^8]$ for YAGO and Wikidata. 
We randomly shuffled all queries and used 80\% of them for training and evaluated all approaches on the remaining 20\%. Prior to this, we removed any duplicates of queries in the dataset (considering isomorphic queries with different variable names).
\begin{table}[h!]
\caption{Queries used for training and testing}
\label{query_counts}
\centering
\footnotesize
\begin{tabular}{|*{9}{c|}}
\hline
KG & \multicolumn{4}{|c}{Star Queries} & \multicolumn{4}{|c|}{Path Queries} \\ 
\hline
\hline 
 \multirow{3}{*}{SWDF} & \multicolumn{4}{|c}{Overall: 116645} & \multicolumn{4}{|c|}{Overall: 60739} \\ \cline{2-9}
& 2tp & 3tp & 5tp & 8tp & 2tp & 3tp & 5tp & 8tp \\  \cline{2-9}
 & 27843 & 29461& 29690 & 29651 & 538 & 19033  & 550 & 40618\\ \hline \hline
 
\multirow{3}{*}{\shortstack{LUBM}} & \multicolumn{4}{|c}{Overall: 113855} & \multicolumn{4}{|c|}{Overall: 55336} \\ \cline{2-9}
& 2tp & 3tp & 5tp & 8tp & 2tp & 3tp & 4tp & 8tp \\  \cline{2-9}
 & 25356 & 29127 & 29727 & 29645 & 17784 & 33047 & 4505 & 0 \\ \hline \hline

 \multirow{3}{*}{YAGO} & \multicolumn{4}{|c}{Overall: 91162} & \multicolumn{4}{|c|}{Overall: 87775}\\ \cline{2-9}
& 2tp & 3tp & 5tp & 8tp & 2tp & 3tp & 5tp & 8tp \\  \cline{2-9}
 &13968 &24416 &24509 &28269 & 19378 & 33807 & 17990 & 16600 \\ \hline \hline
 
  \multirow{3}{*}{Wikidata} & \multicolumn{4}{|c}{Overall: 109545} & \multicolumn{4}{|c|}{Overall: 101313}\\ \cline{2-9}
& 2tp & 3tp & 5tp & 8tp & 2tp & 3tp & 5tp & 8tp \\  \cline{2-9}
 &29975 &29936 &29898 &19736 & 29948 & 29951 & 29997 & 11417 \\ \hline
 
\end{tabular}
\end{table}
\subsubsection{Compared Methods.} 
We compare \textit{GNCE} to the implementation provided by \textit{G-CARE}~\cite{Park2020} of summary-based approaches (\textit{SUMRDF}~\cite{Stefanoni2018} and Characteristic Sets (\textit{CSET})~\cite{charsets}) and sampling\allowbreak -based approaches (\textit{impr}~\cite{Chen2017}, \textit{Wanderjoin}~\cite{Li2016}, and \textit{jsub}~\cite{jsub}), using the recommended default parameters, and refer to the paper~\cite{Park2020} for details. We also compare to state-of-the-art learning-based approaches (\textit{LMKG}~\cite{Davitkova2022} and \textit{LSS}~\cite{lss}). 
For the sampling-based approaches, we report on their average performance over 30 samples (the default setting in \textit{G-CARE)}.
For the learning-based approaches, we trained one model per dataset and query type and set up the number of parameters as recommended by the authors.  
The following table summarizes the number of parameters for the learning-based approaches. 
Note that \textit{GNCE} has the lowest number of parameters, with 18x less w.r.t. \textit{LSS}. 
Fewer parameters typically translate to a smaller memory footprint of the approach and a faster execution time. 
It also indicates that the model is less prone to overfitting. 

\smallbreak
\begin{center}
\footnotesize
\textbf{Number of model parameters\\}
\begin{tabular}{|l|r|r|r|}
\hline
KG & \textit{GNCE} (Ours) & \textit{LMKG} & \textit{LSS} \\
\hline
SWDF & $1.2 \cdot e^5$ & $4.5 \cdot e^5 $ & $2.2 \cdot e^6$ \\
\hline
LUBM & $1.2 \cdot e^5$ & $2.9 \cdot e^5$ & $2.2 \cdot e^6$ \\
\hline
YAGO & $1.2 \cdot e^5$ & $3.8 \cdot e^5$ & $2.2 \cdot e^6$ \\
\hline
Wikidata & $1.2 \cdot e^5$ & $3.1 \cdot e^5$ & $2.2 \cdot e^6$ \\
\hline
\end{tabular}
\end{center}

\subsubsection{Implementation Details}
\textit{GNCE} is implemented using Python 3.8 and Pytorch Geometric. 
For computing the RDF2Vec embeddings, we use the implementation by Vandewiele et al.~\cite{pyrdf2vec}.
We train the RDF2Vec model for 10 epochs. For each entity, we sample 5 random walks with a maximum depth of 4 and learn the embeddings as presented in Section~\ref{sec:embeddings}. The dimensionality of the embeddings is set to 100. Finally, we store the embeddings for later use by \textit{GNCE}.
All experiments, except for \textit{LSS} on YAGO, have been conducted on a machine with 16 GB RAM, Intel Core i7-11800H@2.30GHz, and an NVIDIA GeForce RTX 3050 GPU. 
Since \textit{LSS} requires extensive amounts of RAM as it loads the whole KG into memory for training, it was trained on a machine with 512 GB RAM, an Intel Xeon Silver 4314 CPU@2.40GHz, and an Nvidia A100 GPU.

\subsubsection{Evaluation Metric}
Similar to previous works~\cite{lss}, we evaluate the effectiveness of the approaches in terms of the q-Error:
\begin{equation}
\text{q-Error}(Q) = \max \left( \frac{||Q||_\mathcal{G}}{\widehat{||Q||}_\mathcal{G}}, \frac{\widehat{||Q||}_\mathcal{G}}{||Q||_\mathcal{G}} \right)
\end{equation}

\subsection{Cardinality Estimation for Star Queries}
\label{sec:star}

In this section, we study the effectiveness of the approaches for star queries, where the subject position is the join variable. 

In Table~\ref{q_table_star}, we report the mean q-Error of all approaches in each dataset.\footnote{Results for \textit{SUMRDF} on YAGO and Wikidata could not be reported, as the summary building for that graph did not finish after 12 hours.}  
Overall, we observe that the q-Error increases along with the size of the KGs.

This behavior is expected, as capturing correlations in larger datasets is more challenging, and queries with large cardinalities naturally occur more frequently in these datasets. 
These aspects negatively impact the estimations.  
The sampling and summary methods behave similarly to some extent, except for \textit{jsub} which produces large errors in YAGO.  
\textit{CSET} does not perform well in the star queries tested, as these may have bound objects, which can create overestimations in \textit{CSET}.  

Among the learned approaches, \textit{GNCE} exhibits the best overall performance.      
Interestingly, \textit{LSS} does not perform well in these KGs. 
We hypothesize that this is because the learned node representations based on classes of \textit{LSS} cannot distinguish between queries with similar shapes but with different cardinalities. This behaviour of \textit{LSS} is clearly observed in SWDF. 
While all other approaches achieve their best performance in SWDF (smallest dataset), \textit{LSS} produces large q-Errors because the number of typed entities is considerably small (only 22\%, cf. Table~\ref{tab:datasets}). 
In the case of untyped entities, the \textit{LSS} entity representation defaults to a generic vector of zeros, which does not contain any information. 
In contrast, \textit{GNCE} does not make any assumption about the types of entities, and can still encode meaningful representations of the entities through the KG embeddings.

\begin{figure}[!]
\centering     
\subfigure[SWDF]{\label{fig:b}\includegraphics[width=0.24\textwidth]{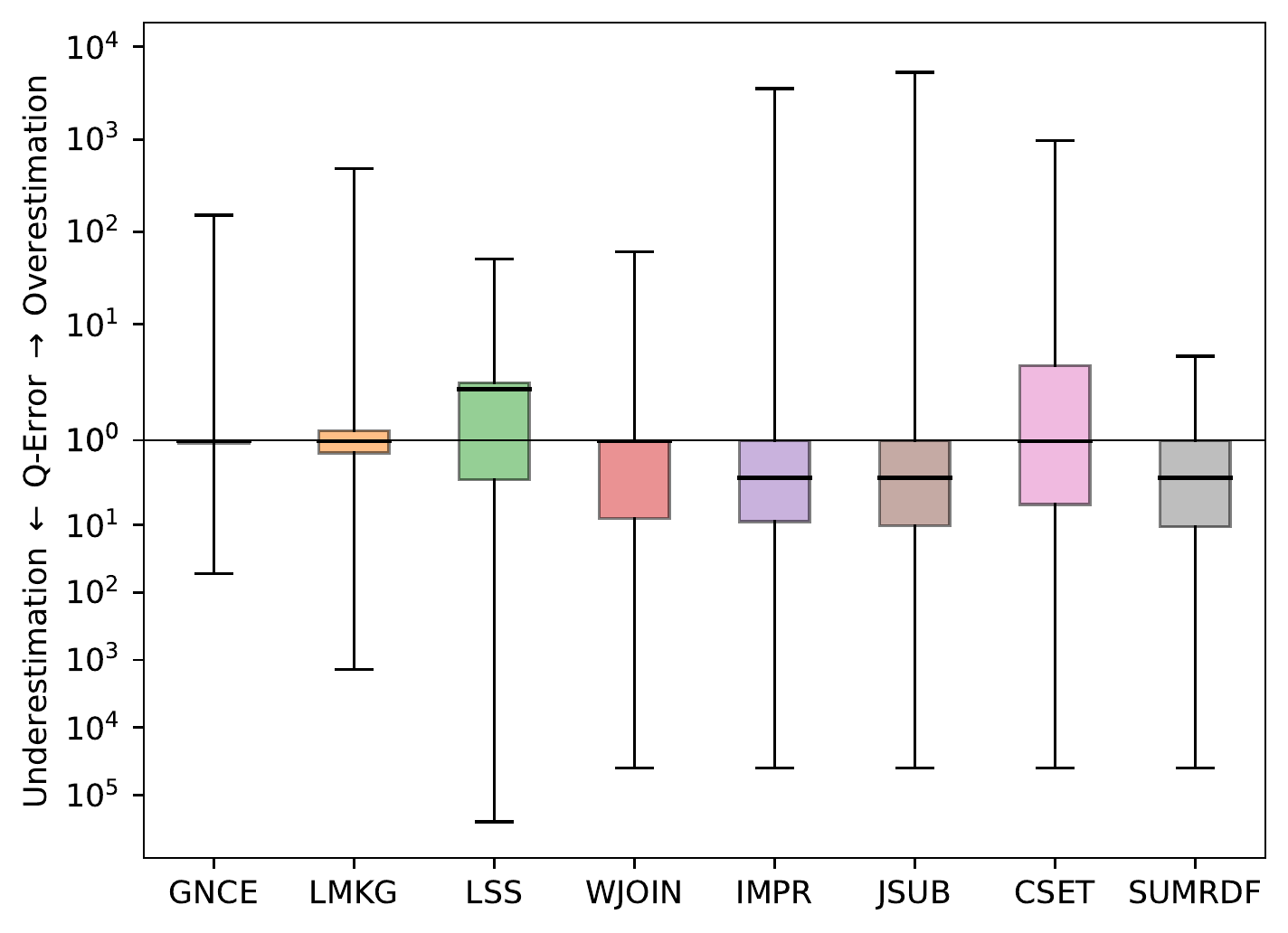}}
\subfigure[LUBM]{\label{fig:a}\includegraphics[width=0.24\textwidth]{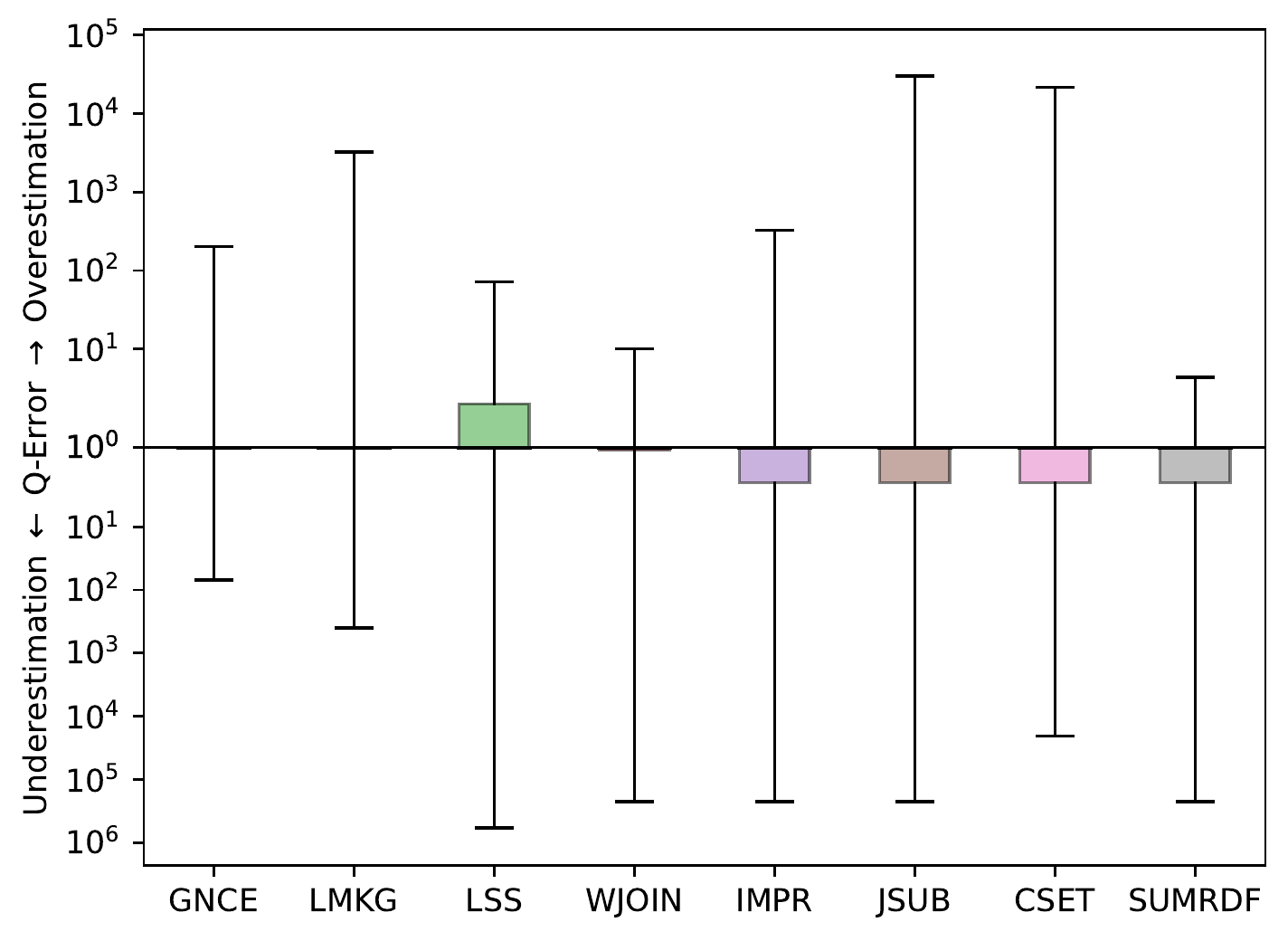}}
\subfigure[YAGO]{\label{fig:c}\includegraphics[width=0.24\textwidth]{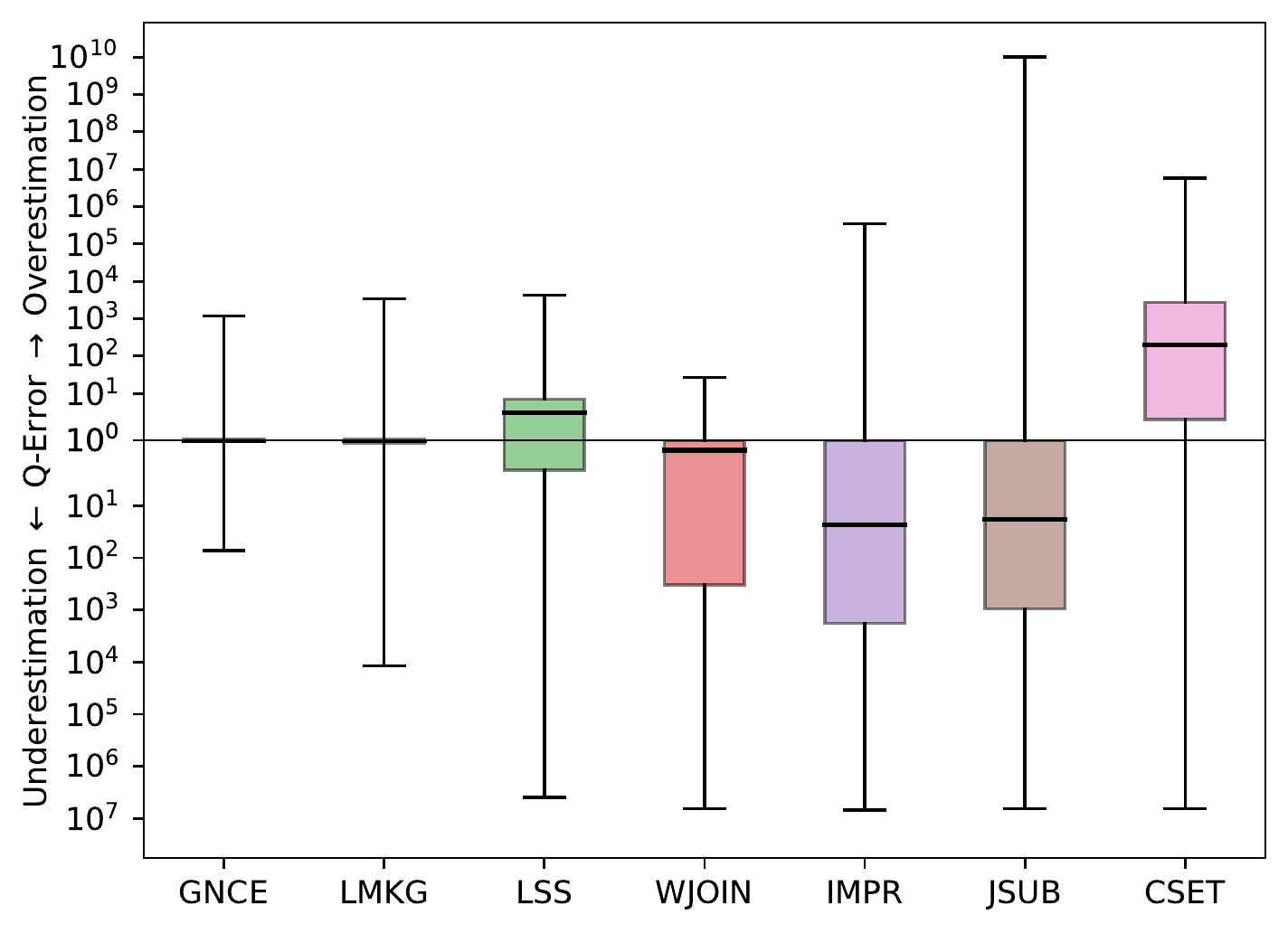}}
\subfigure[Wikidata]{\label{fig:d}\includegraphics[width=0.24\textwidth]{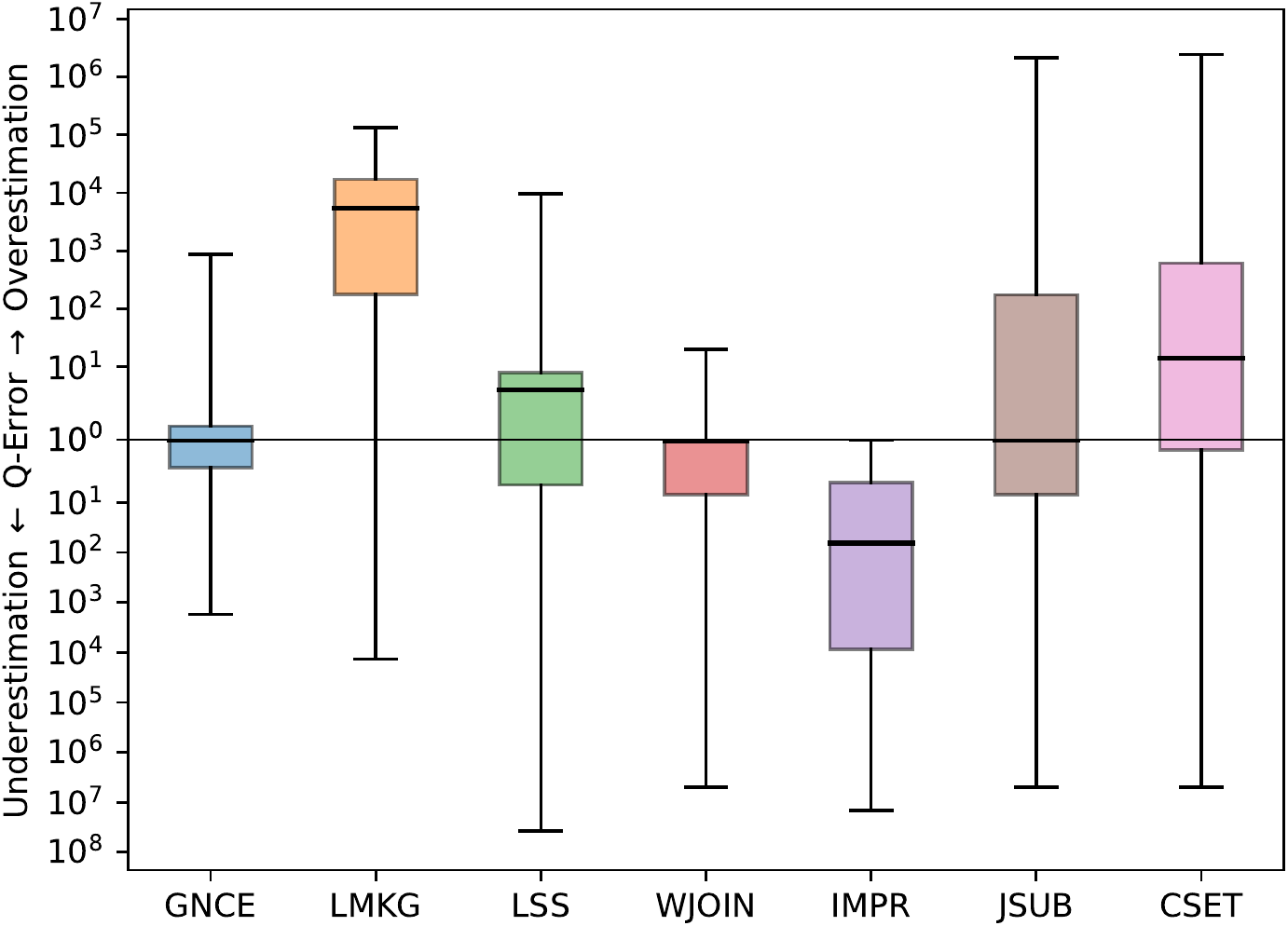}}
\caption{Boxplots of q-Errors of star queries}
\label{fig:boxplot_star}
\end{figure}
\begin{figure*}[t!]
\centering     
\subfigure[SWDF]{\label{fig:bar_star_swdf}\includegraphics[width=0.24\textwidth]{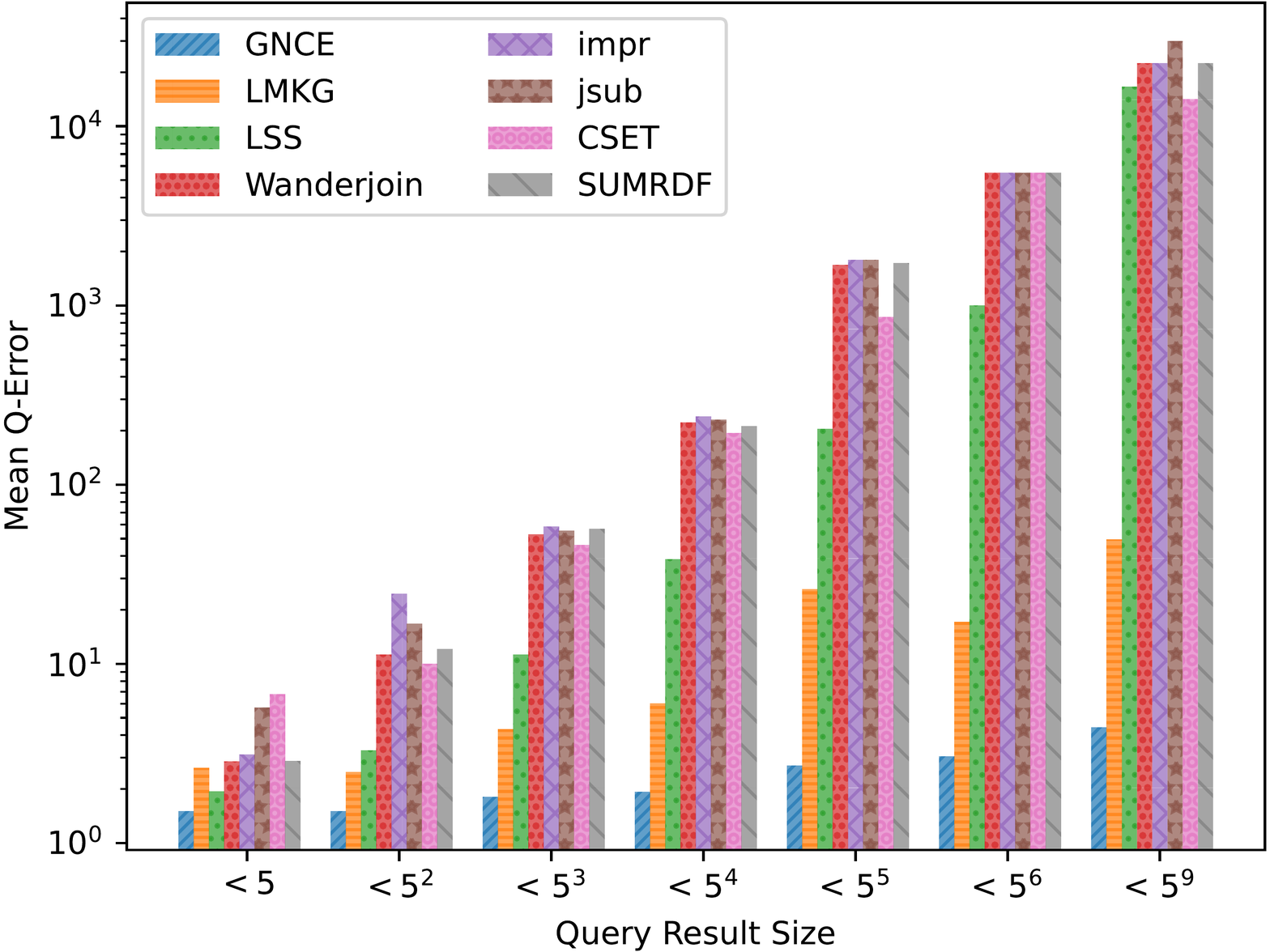}}
\subfigure[LUBM]{\label{fig:bar_star_lubm}\includegraphics[width=0.24\textwidth]{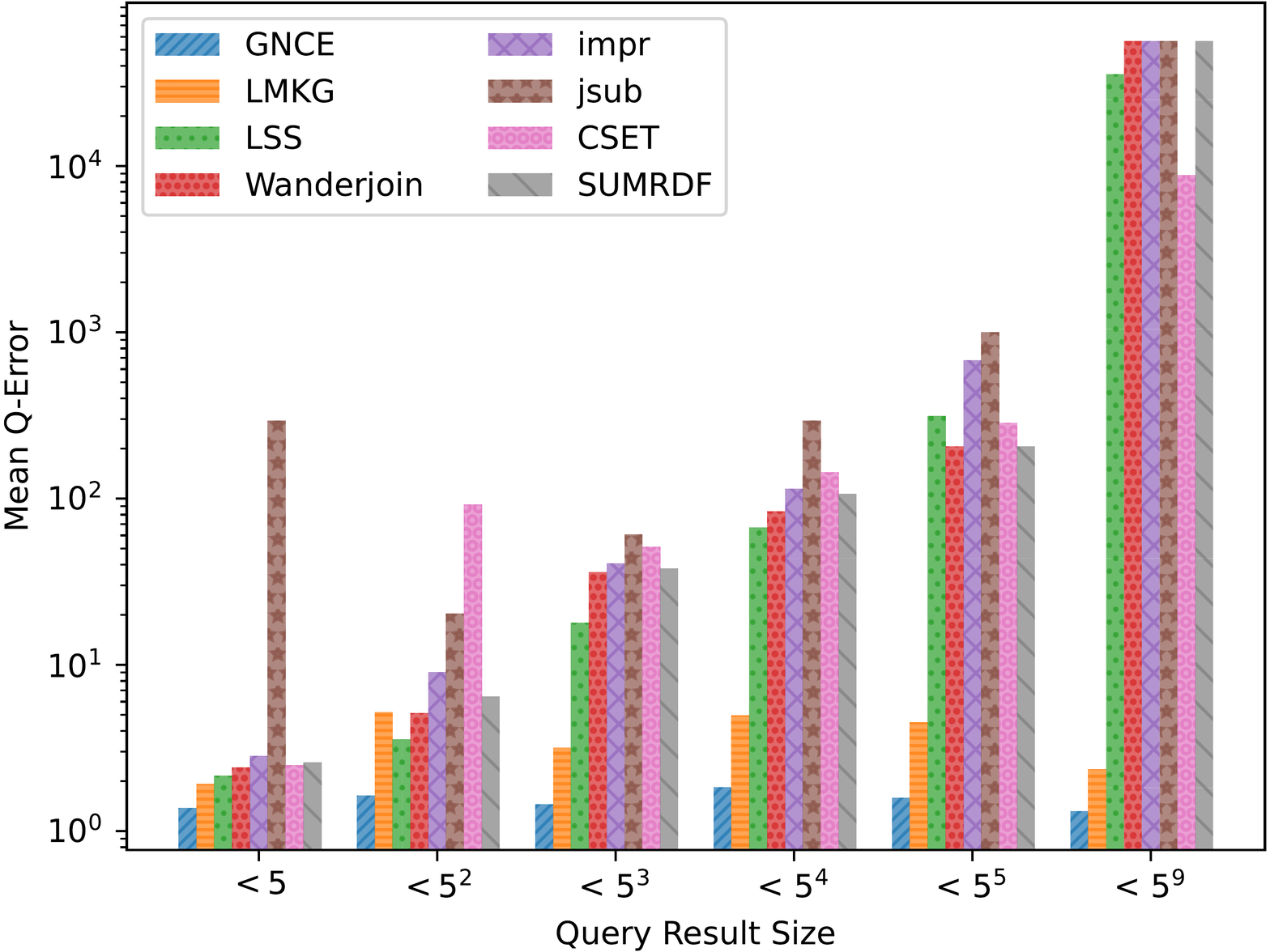}}
\subfigure[YAGO]{\label{fig:bar_star_yago}\includegraphics[width=0.24\textwidth]{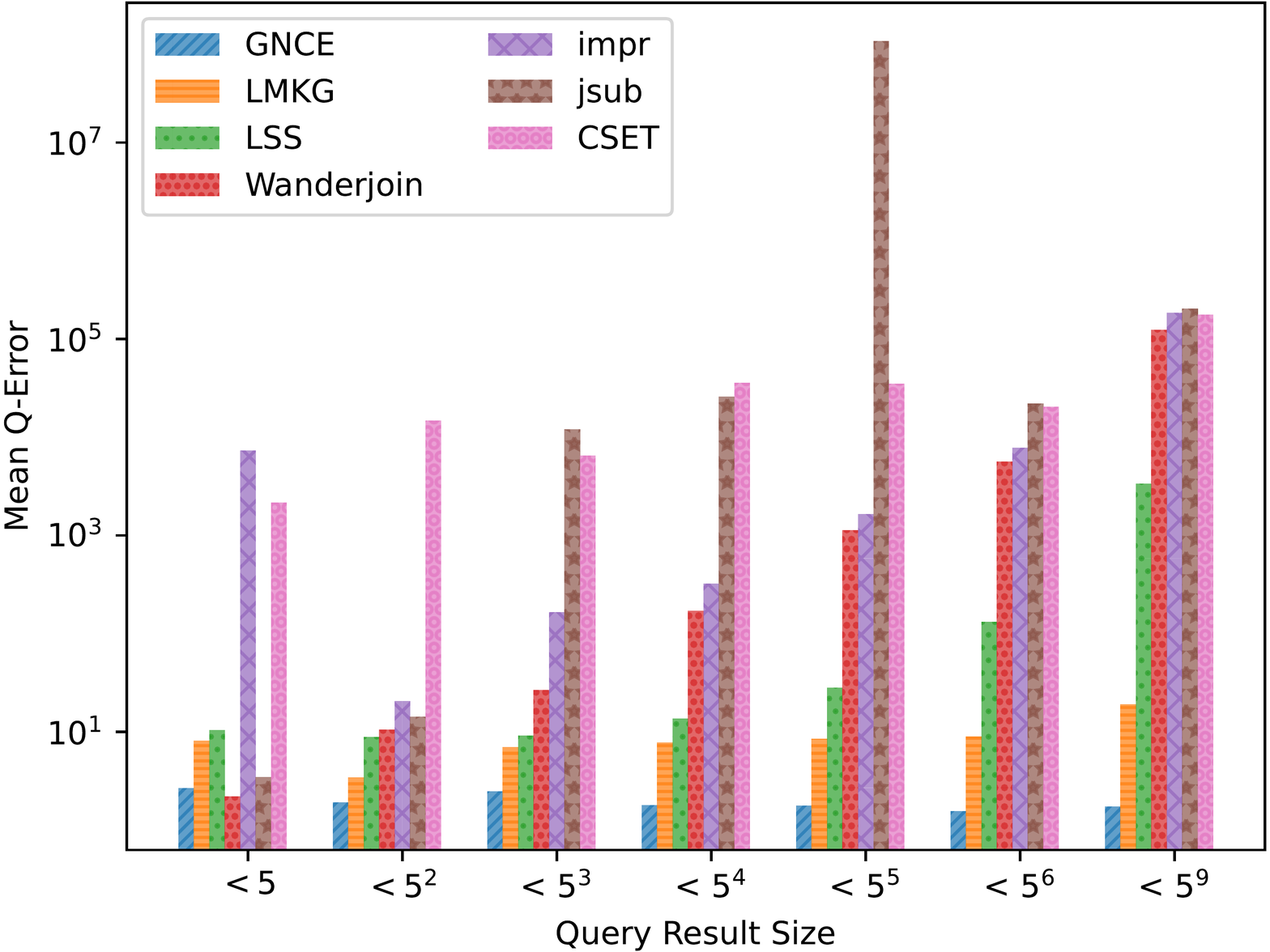}}
\subfigure[Wikidata]{\label{fig:bar_star_wikidata}\includegraphics[width=0.24\textwidth]{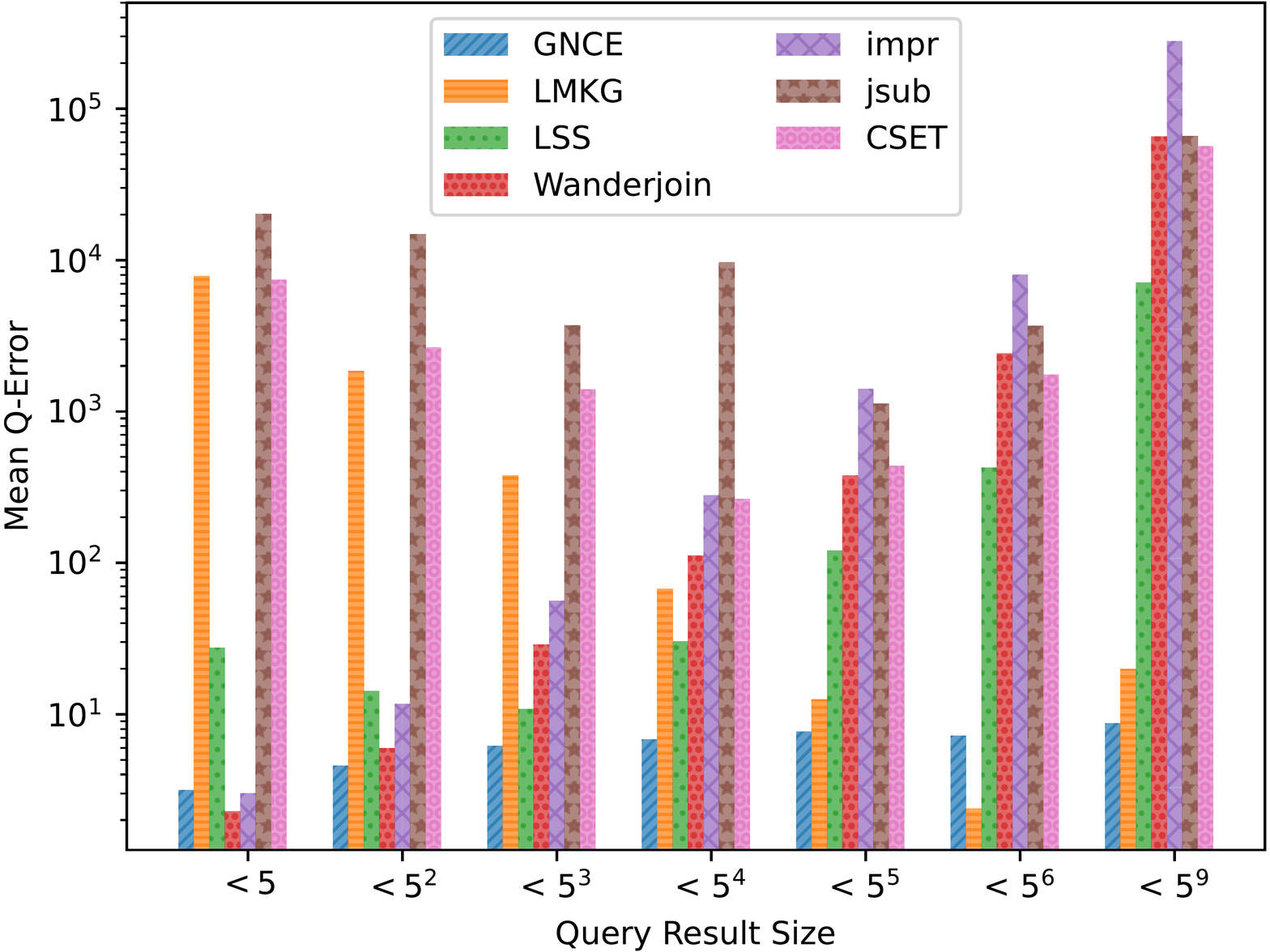}}
\caption{q-Errors for star-shaped queries, grouped by the true cardinalities of the queries}
\label{fig:barplot_star}
\end{figure*}
\begin{figure*}[t!]
\centering     
\subfigure[SWDF]{\label{fig:b}\includegraphics[width=0.24\textwidth]{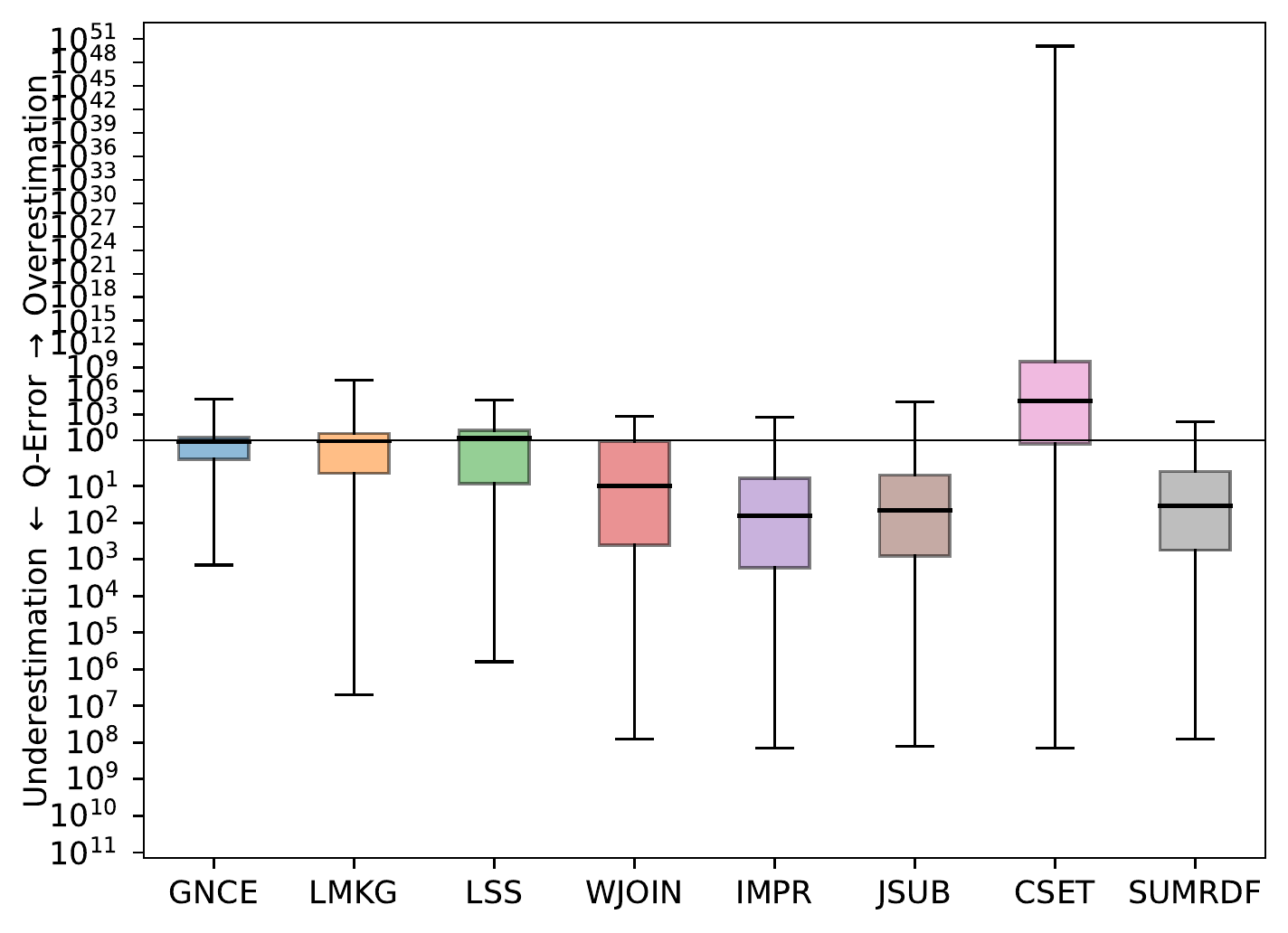}}
\subfigure[LUBM]{\label{fig:a}\includegraphics[width=0.24\textwidth]{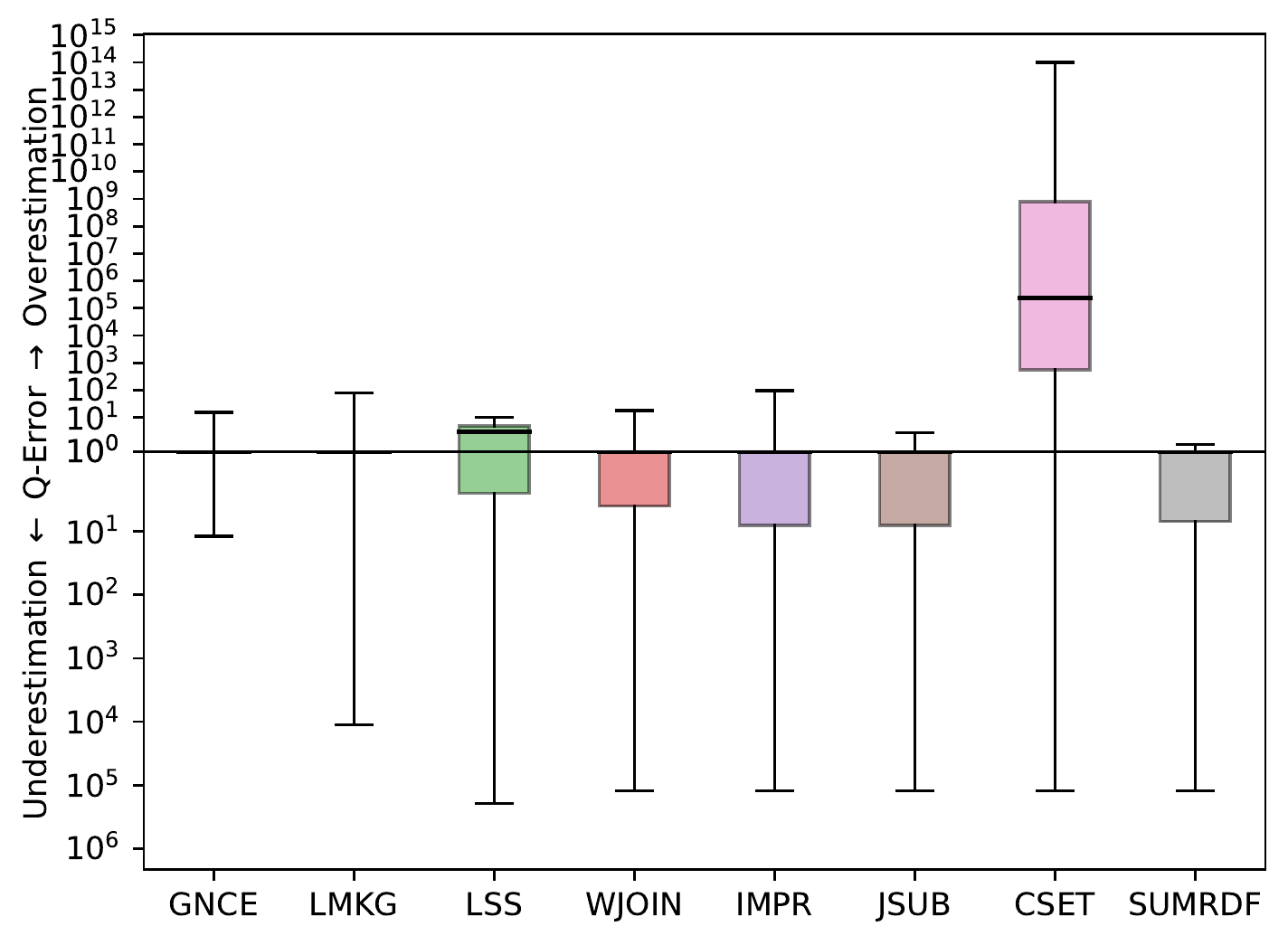}}
\subfigure[YAGO]{\label{fig:c}\includegraphics[width=0.24\textwidth]{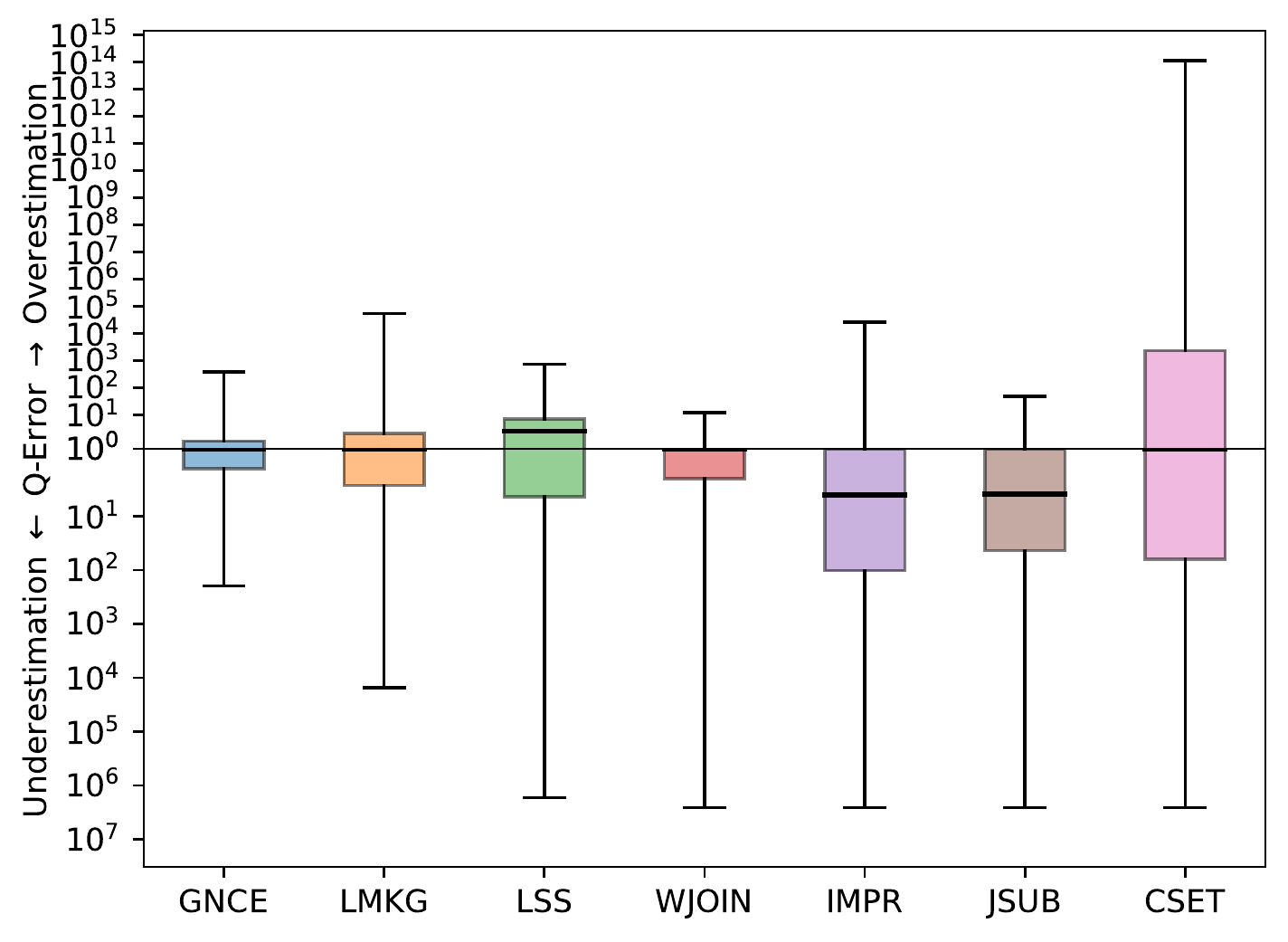}}
\subfigure[Wikidata]{\label{fig:c}\includegraphics[width=0.24\textwidth]{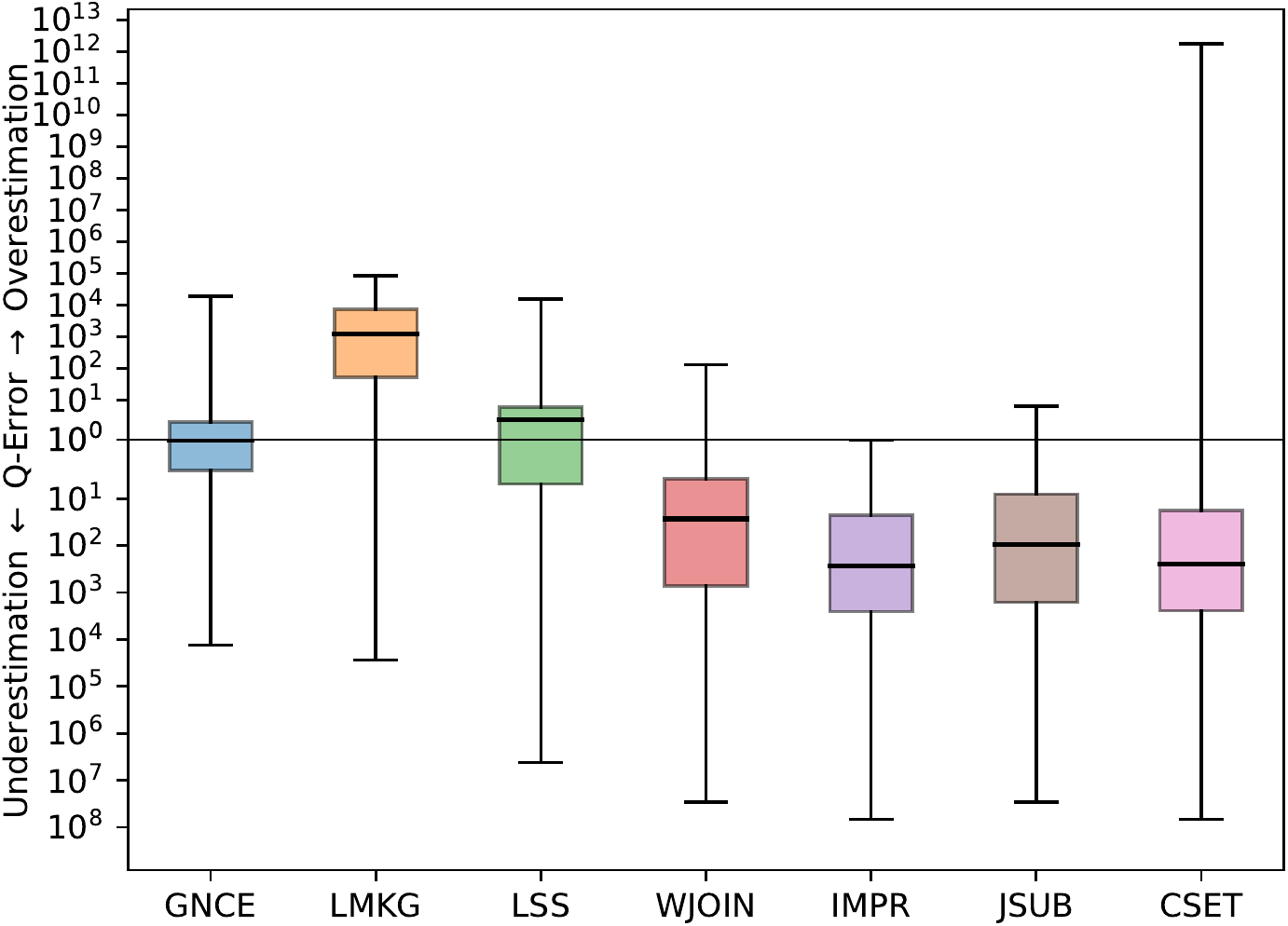}}
\caption{Boxplots of q-Errors for path queries}
\label{fig:boxplot_path}
\end{figure*}
\begin{figure*}[t!]
\centering     
\subfigure[SWDF]{\label{fig:path_b}\includegraphics[width=0.24\textwidth]{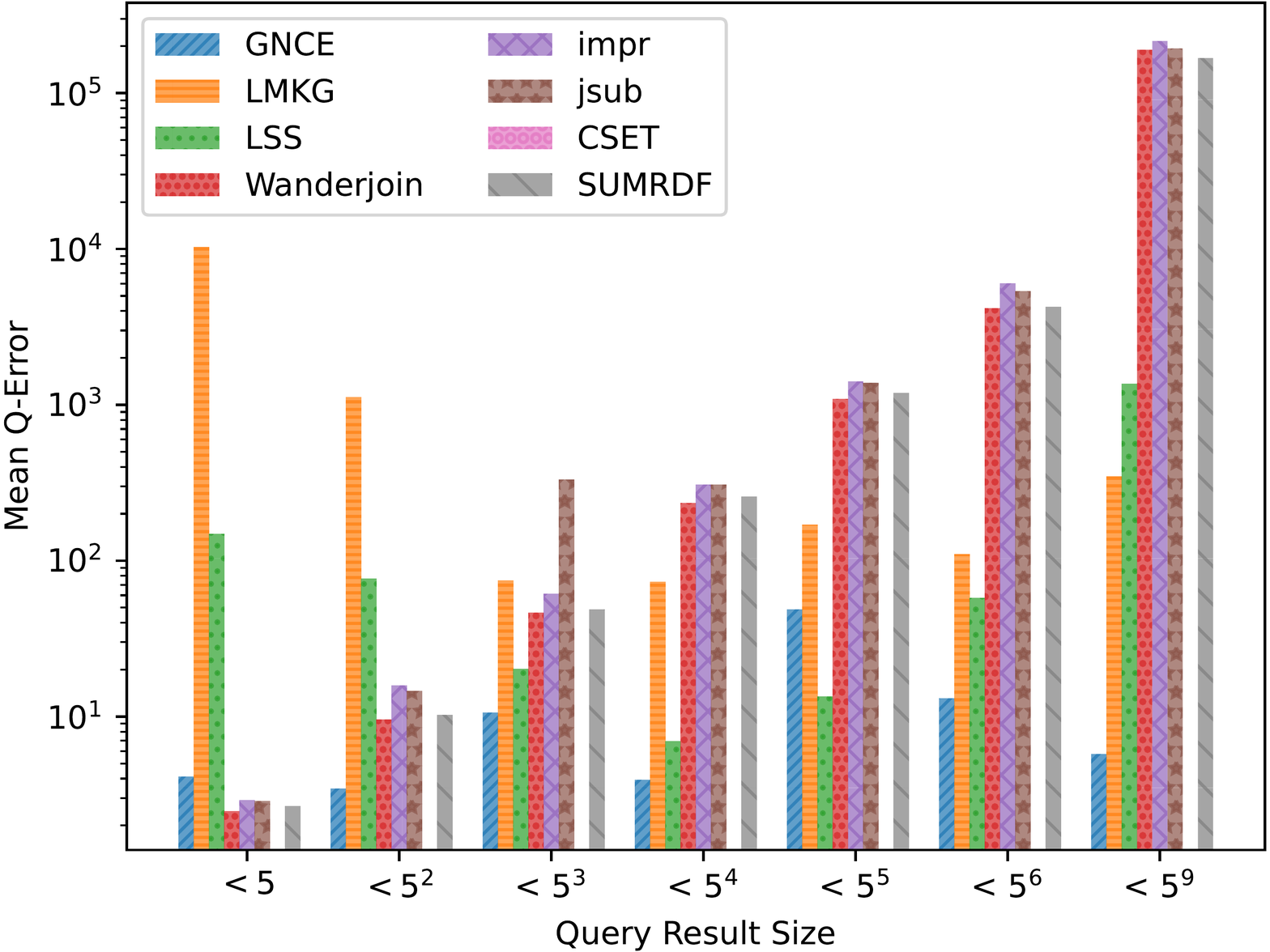}}
\subfigure[LUBM]{\label{fig:path_a}\includegraphics[width=0.24\textwidth]{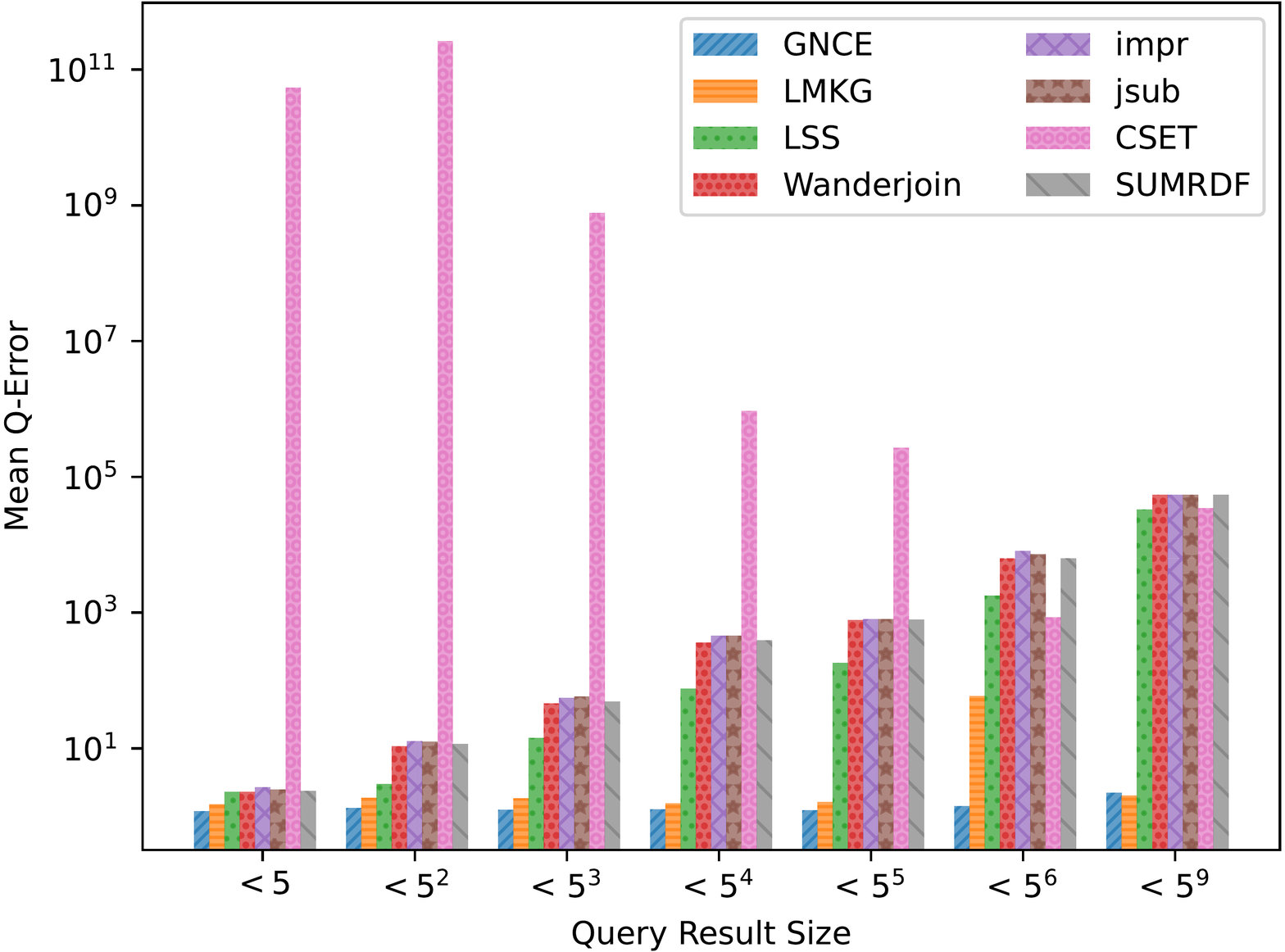}}
\subfigure[YAGO]{\label{fig:path_c}\includegraphics[width=0.24\textwidth]{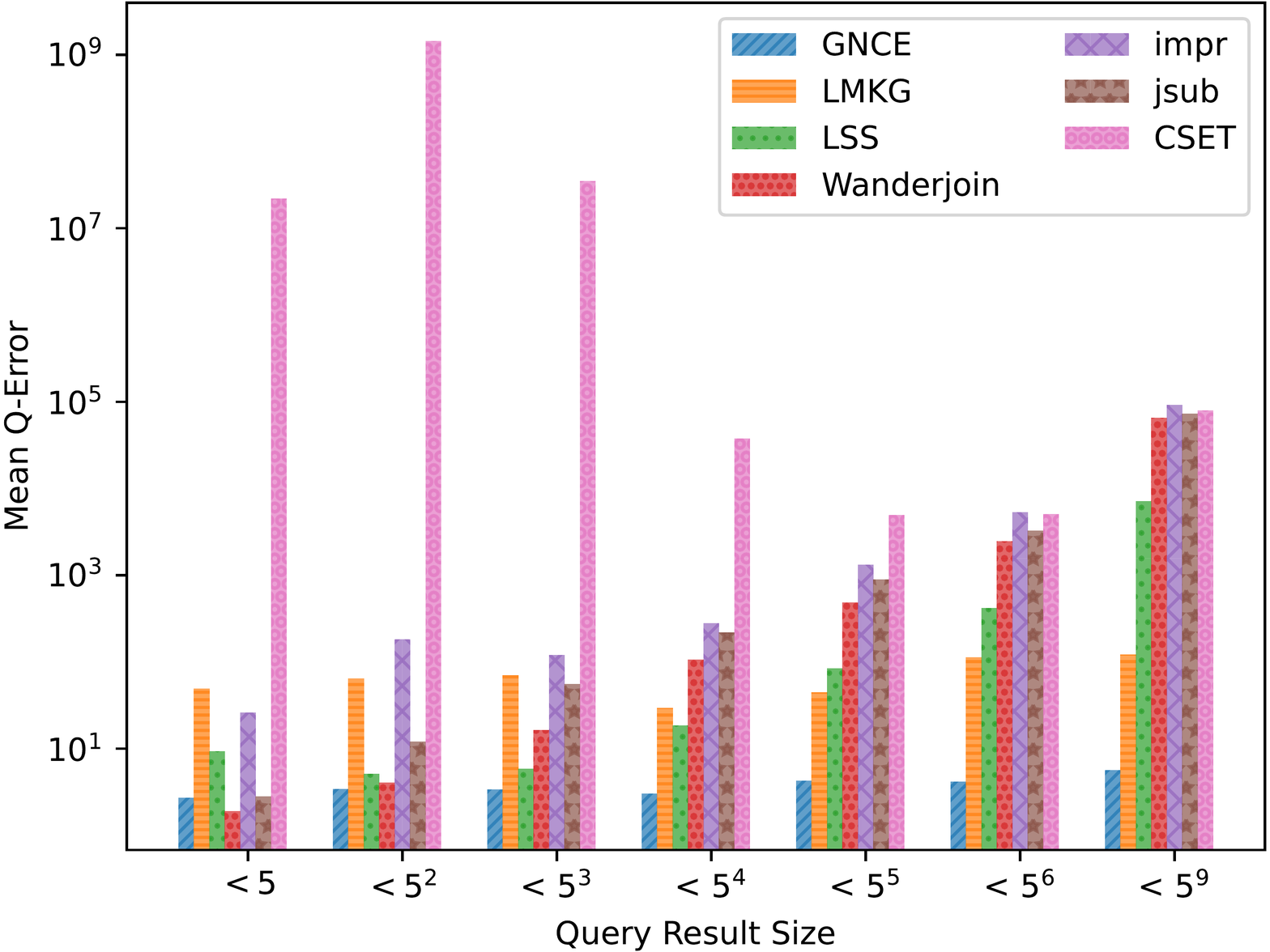}}
\subfigure[Wikidata]{\label{fig:path_c}\includegraphics[width=0.24\textwidth]{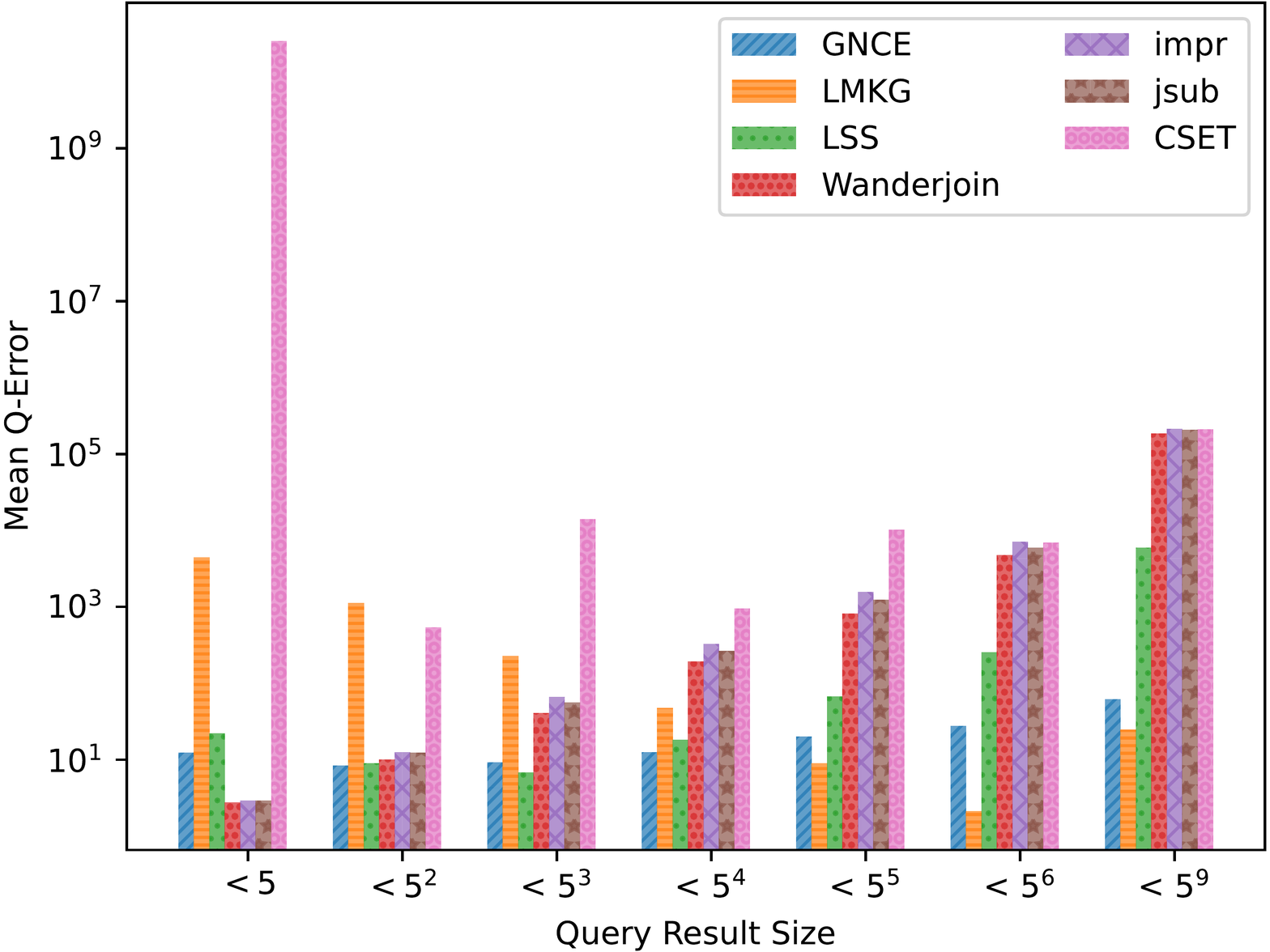}}
\caption{q-Errors for path queries, grouped by the true cardinalities of the queries}
\label{fig:barplot_path}
\end{figure*}

\begin{table}[t!]
\caption{Mean q-Errors for star queries}
\label{q_table_star}
\centering
\footnotesize
\begin{tabular}{r|r|r|r|r|r|}
\cline{2-6}
\multicolumn{1}{r|}{}                           & \textbf{Approach} & \textbf{SWDF} & \textbf{LUBM} & \textbf{YAGO} & \textbf{Wikidata}    \\ \hline
\multicolumn{1}{|l|}{\multirow{3}{*}{Sampling}} & \textit{Wanderjoin}        & 129.00           & 232.00           & 18713.00    & 34564.00        \\ \cline{2-6} 
\multicolumn{1}{|l|}{}                          & \textit{impr}              & 141.00           & 237.00           & 2605.00    & 169241.00         \\ \cline{2-6} 
\multicolumn{1}{|l|}{}                          & \textit{jsub}              & 169.00           & 409.00           & 15 $\cdot e^{6}$ & 40055.00
\\ \hline
\multicolumn{1}{|l|}{\multirow{2}{*}{Summary}}  & \textit{CSET}              & 94.00            & 904.00           & 41586.00    & 37310.00        \\ \cline{2-6} 
\multicolumn{1}{|l|}{}                          & \textit{SUMRDF}            & 130.00           & 233.00           & -    & -           \\ \hline
\multicolumn{1}{|l|}{\multirow{3}{*}{Learned}}  & \textit{LMKG}              & 3.00           & 2.00          & 9.00   & 10267.00          \\ \cline{2-6} 
\multicolumn{1}{|l|}{}                          & \textit{LSS}               & 1187.00          & 283.00           & 502.00    & 4089.00        \\ \cline{2-6} 
\multicolumn{1}{|l|}{}                          & \textit{GNCE}              & 
\textbf{1.42} & \textbf{1.22} & \textbf{1.96} & \textbf{4.20}    \\ \cline{1-6} 
\end{tabular}
\vspace{-2mm}
\end{table}

To further understand the performance of the approaches, we analyze the distributions of q-Errors. Figure~\ref{fig:boxplot_star} shows the distribution of q-Errors as boxplots;  under- and overestimations are depicted by values below and above the horizontal line (perfect q-Error), respectively.
We start by looking at the sampling methods (\textit{Wanderjoin}, \textit{impr}, and \textit{jsub}), which tend to underestimate the cardinalities, in particular \textit{impr}. 
This behavior was already observed by Park et al.~\cite{Park2020} for other datasets. 
The summary-based approaches also tend to produce underestimations, except for \textit{CSET} in YAGO and Wikidata. 
A closer inspection of the results reveals that \textit{CSET} largely overestimates the cardinalities for highly selective queries with bound objects. 
For the learned-based methods, the q-Errors are concentrated around the optimal value of 1 for \textit{GNCE} and \textit{LMKG}. 
\textit{LSS} tends to overestimate slightly on large parts of the queries, but also shows severe underestimation. 
In comparison to \textit{LSS} and \textit{LMKG}, the maximum under-and overestimations of \textit{GNCE}, as well as the quantiles, are more symmetric around the origin than for the other methods, indicating a more stable behavior across all datasets. 


Next, we analyze the average q-Error of the approaches at different cardinality sizes (cf. Figure~\ref{fig:barplot_star}).   
The results show that the q-Error tends to grow with increasing query cardinality, except for \textit{GNCE} and \textit{LMKG}. 
Non-learning-based approaches exhibit good performance in highly selective queries with a result size range < 5. 
These queries typically have many bound atoms, for which sampling-based methods like \textit{impr} and \textit{Wanderjoin} might have collected (nearly) exact statistics. 
The only exception is \textit{CSET} in YAGO, yet, these results are due to the large q-Errors produced by overestimations of \textit{CSET} as previously explained. 
For learning-based approaches, the low number of training queries in the higher cardinality range can contribute to large errors. 
Still, \textit{GNCE}, with its inductive bias tailored to the graph nature of queries and more informative feature representation due to embeddings, requires less training data to produce accurate predictions.
For this reason, \textit{GNCE} consistently has the lowest q-Error across all result sizes, with the only exception being the result size range < 5 in YAGO and Wikidata, where it is slightly outperformed by \textit{Wanderjoin}. 

\subsection{Cardinality Estimation for Path Queries}
\label{sec:path}

Next, we examine the effectiveness of the approaches for path queries, with joins occurring between subjects and objects of triple patterns.  
Table~\ref{q_table_path} presents the mean q-Errors. 
Most methods have higher q-Errors in comparison to the star-shaped queries (cf.~Table~\ref{q_table_star}). 
This is consistent with previous results reported by Park et al.~\cite{Park2020}.  
Also interestingly, all methods are at least an order of magnitude worse on SWDF compared to the other datasets. 
This can be due to the high density of the SWDF graph, i.e., there are lots of possible paths from each starting point which makes cardinality estimation significantly more error-prone. 
The effectiveness of the learning-based approaches is also affected by the fewer (training) queries in the dataset (cf. Table~\ref{query_counts}).\footnote{Please note that this is not a problem of the query generator, but rather that KGs in general exhibit a small diameter~\cite{DBLP:conf/esws/ZlochAHD019}, which makes the number of different path queries considerable smaller than the number of star queries.} 
Overall, the learned approaches exhibit the best performance, followed by sampling-based methods, and lastly summary-based methods. 
Still, \textit{GNCE} outperforms the state of the art in path queries by several orders of magnitude.

\begin{table}[t!]
\caption{Mean q-Errors for path queries}
\label{q_table_path}
\centering
\footnotesize
\begin{tabular}{r|r|r|r|r|r|}
\cline{2-6}
                                                & \textbf{Approach}   & \textbf{SWDF}      & \textbf{LUBM}     & \textbf{YAGO}   
                                                & \textbf{Wikidata}\\ \hline
\multicolumn{1}{|l|}{\multirow{3}{*}{Sampling}} & \textit{Wanderjoin} & 22 $\cdot e^4$   & 389.00             & 9730.00     & 67001.00          \\ \cline{2-6} 
\multicolumn{1}{|l|}{}                          & \textit{impr}       & $73.7 \cdot e^4$   & 429.00             & 1722.00      & 146623.00         \\ \cline{2-6} 
\multicolumn{1}{|l|}{}                          & \textit{jsub}       & $37.4 \cdot e^4$   & 414.00            & 10952.00      & 69822.00        \\ \hline
\multicolumn{1}{|l|}{\multirow{2}{*}{Summary}}  & \textit{CSET}       & $1 \cdot e^{47}$ & $63 \cdot e^{10}$ & $2 \cdot e^{29}$
& $9 \cdot e^{30}$\\ \cline{2-6} 
\multicolumn{1}{|l|}{}                          & \textit{SUMRDF}     & $2 \cdot e^{5}$    & $393 \cdot e^6$   & -  & -                \\ \hline
\multicolumn{1}{|l|}{\multirow{3}{*}{Learned}}  & \textit{LMKG}       & 1820.00             & 2.00              & 69.00    & 4220.00             \\ \cline{2-6} 
\multicolumn{1}{|l|}{}                          & \textit{LSS}        & 1187.00             & 179.00               & 1075.00   & 1204.00            \\ \cline{2-6} 
\multicolumn{1}{|l|}{}                          & \textit{GNCE}       & \textbf{12.50}      & \textbf{1.14}     & \textbf{3.78} & \textbf{14.00}     \\ \cline{1-6} 
\end{tabular}
\vspace{-2mm}
\end{table}

Next, we analyze the distribution of q-Errors in Figure~\ref{fig:boxplot_path}.  
All the sampling methods tend to underestimate the query cardinalities, especially in SWDF as previously explained due to the density of this dataset. 
Furthermore, their performance greatly deteriorates for YAGO and Wikidata (in comparison to LUBM) due to the size and the known semi-structuredness of these KGs.  
The estimates computed by \textit{CSET} are unrealistically high, a phenomenon that was already reported by Davitkova et al.~\cite{Davitkova2022}.
These results confirm that \textit{CSET} summaries do not perform well at capturing join correlations between subjects and objects, as expected. 
In the learning-based approaches, \textit{LSS} tends to produce overestimations. 
This was already observed in the star queries (cf. Section~\ref{sec:star}), indicating that this is a recurrent behavior of \textit{LSS}. 
In contrast, both \textit{GNCE} and \textit{LMKG} tend to underestimate, except for \textit{LMKG} on Wikidata. Still, the spread of the errors shown by the whiskers is smaller for \textit{GNCE} than \textit{LMKG}.

In Figure~\ref{fig:barplot_path}, we examine the q-Errors categorized by query cardinality.
Figure~\ref{fig:path_b} omits the bars for \textit{CSET}  due to the extremely high values.
Similar to the trend observed for star-shaped queries, we found that the q-Error increases with the cardinality of queries for all approaches except for \textit{GNCE}, which exhibited a mostly consistent mean q-Error.
Second, in SWDF, YAGO, and Wikidata, the sampling-based methods marginally outperform \textit{GNCE} in the small cardinality range <5. Interestingly, on SWDF, for queries with cardinalities between $5^4$ and $5^5$, \textit{LSS} performs best out of all approaches in this particular case. Lastly, \textit{GNCE} performs better than \textit{LMKG} in the smaller cardinality range on Wikidata, while \textit{LMKG} is better in the high cardinality range.
\subsection{Cardinality Estimation for User and Complex Queries}
\label{sec:complex}
To analyze the approaches' effectiveness beyond synthetic star- and path queries, we also investigate more challenging types of queries, namely real user queries and more complex query shapes.

\begin{figure*}[h!]
\subfigure[Wikidata User Queries]{\label{fig:b}\includegraphics[width=0.45\textwidth]{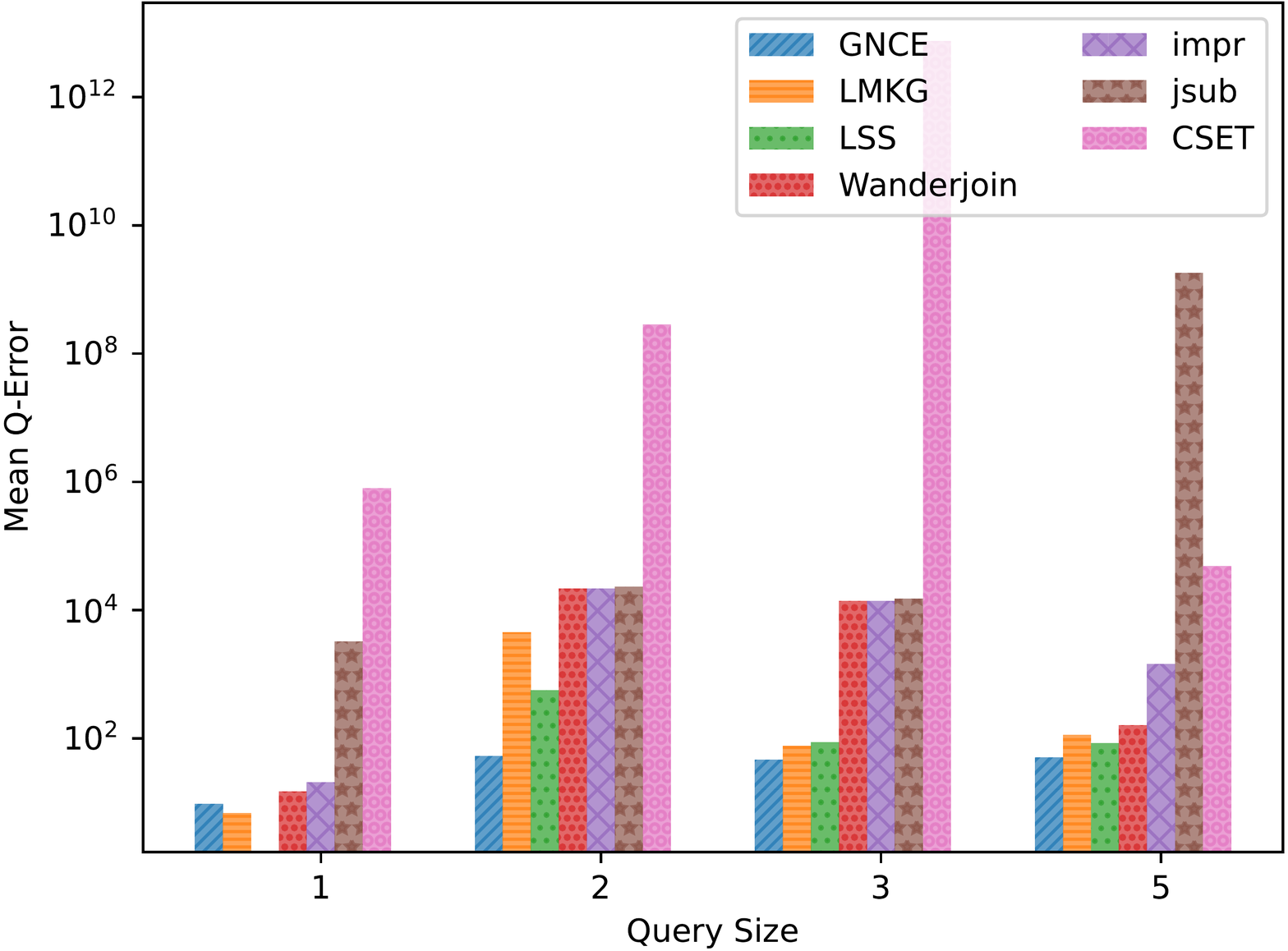}}
\hfill
\subfigure[Wikidata User Queries]{\label{fig:a}\includegraphics[width=0.45\textwidth]{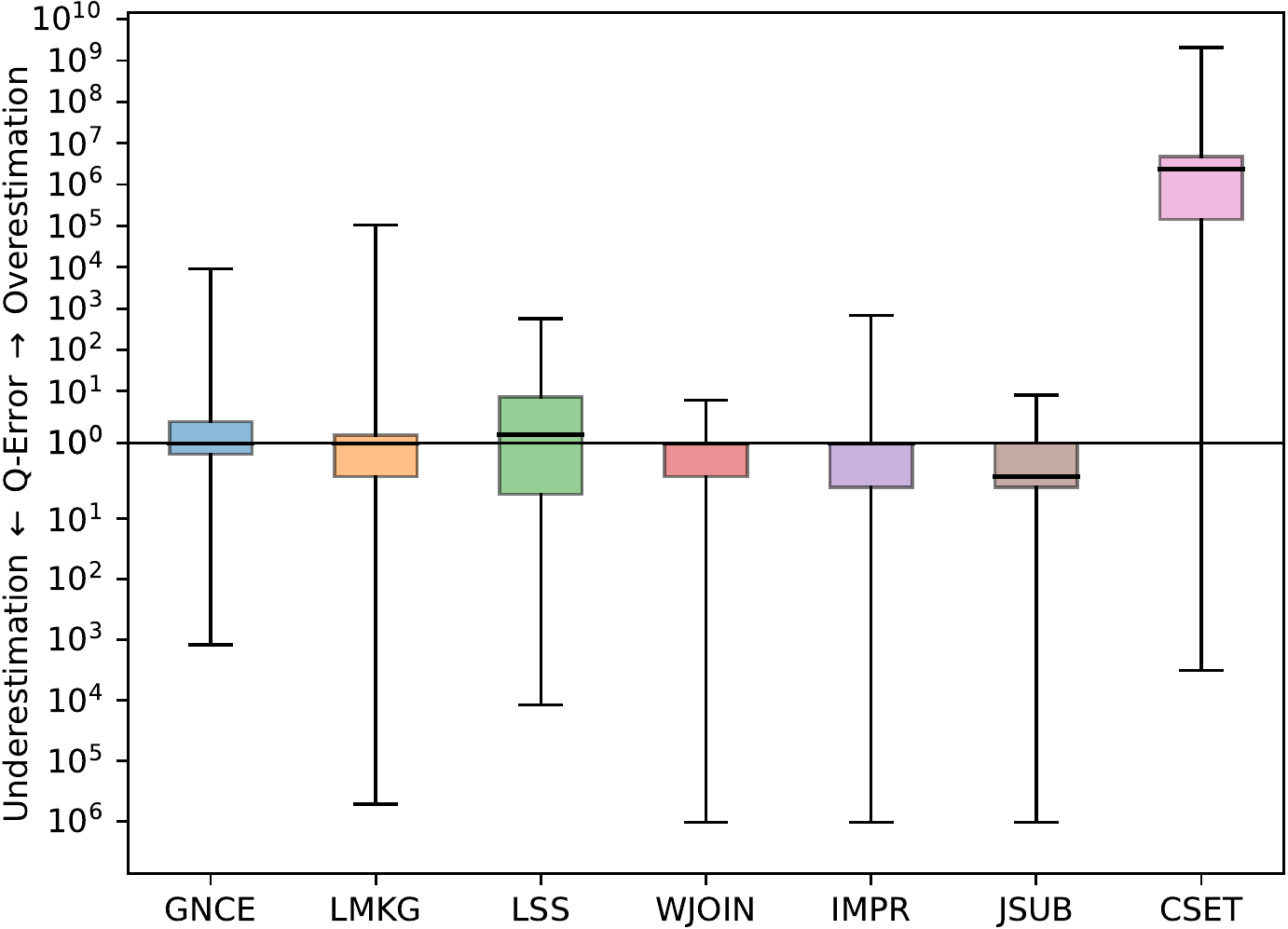}}

\subfigure[YAGO Complex]{\label{fig:b}\includegraphics[width=0.45\textwidth]{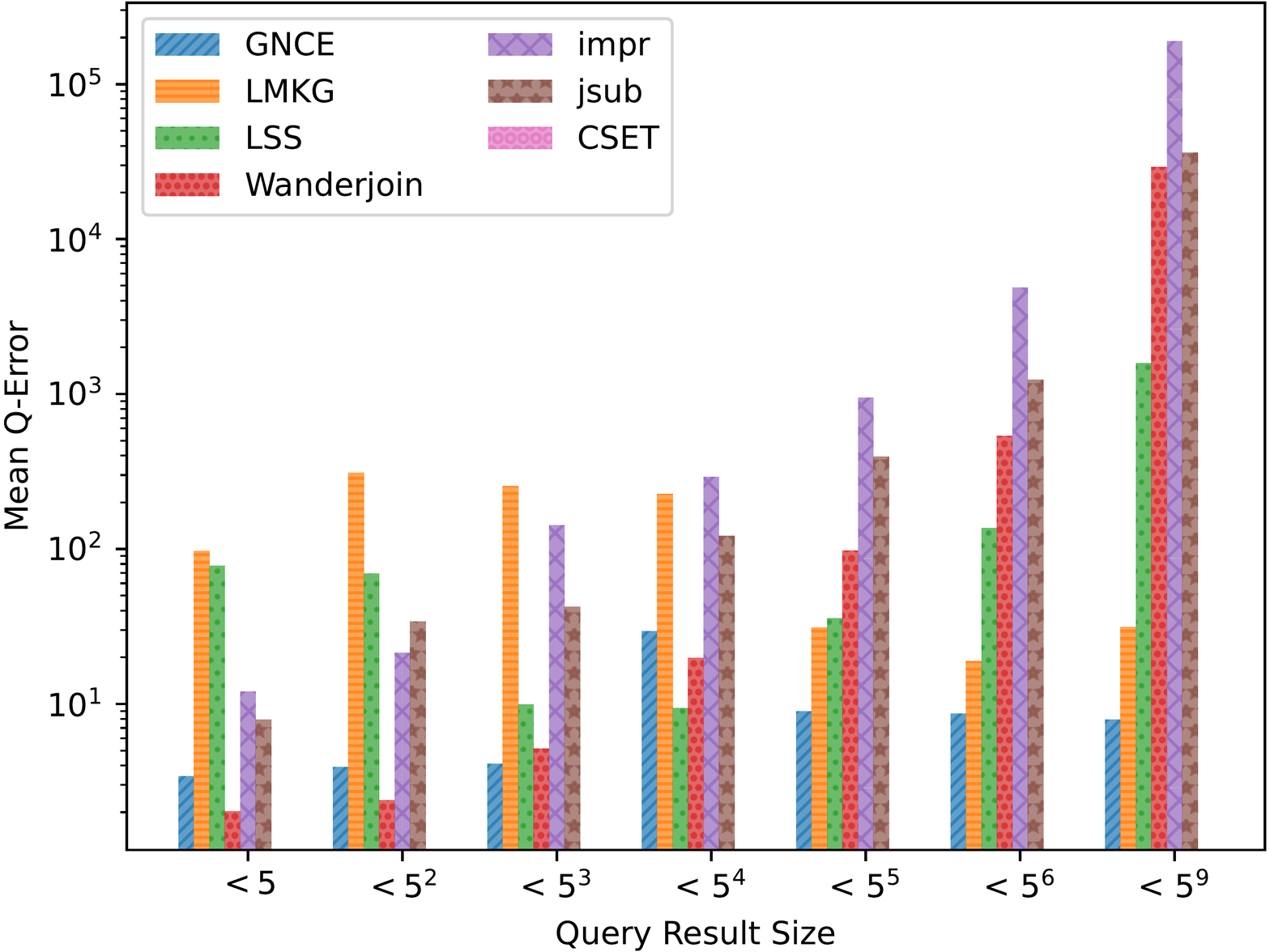}}
\hfill
\subfigure[YAGO Complex]{\label{fig:a}\includegraphics[width=0.45\textwidth]{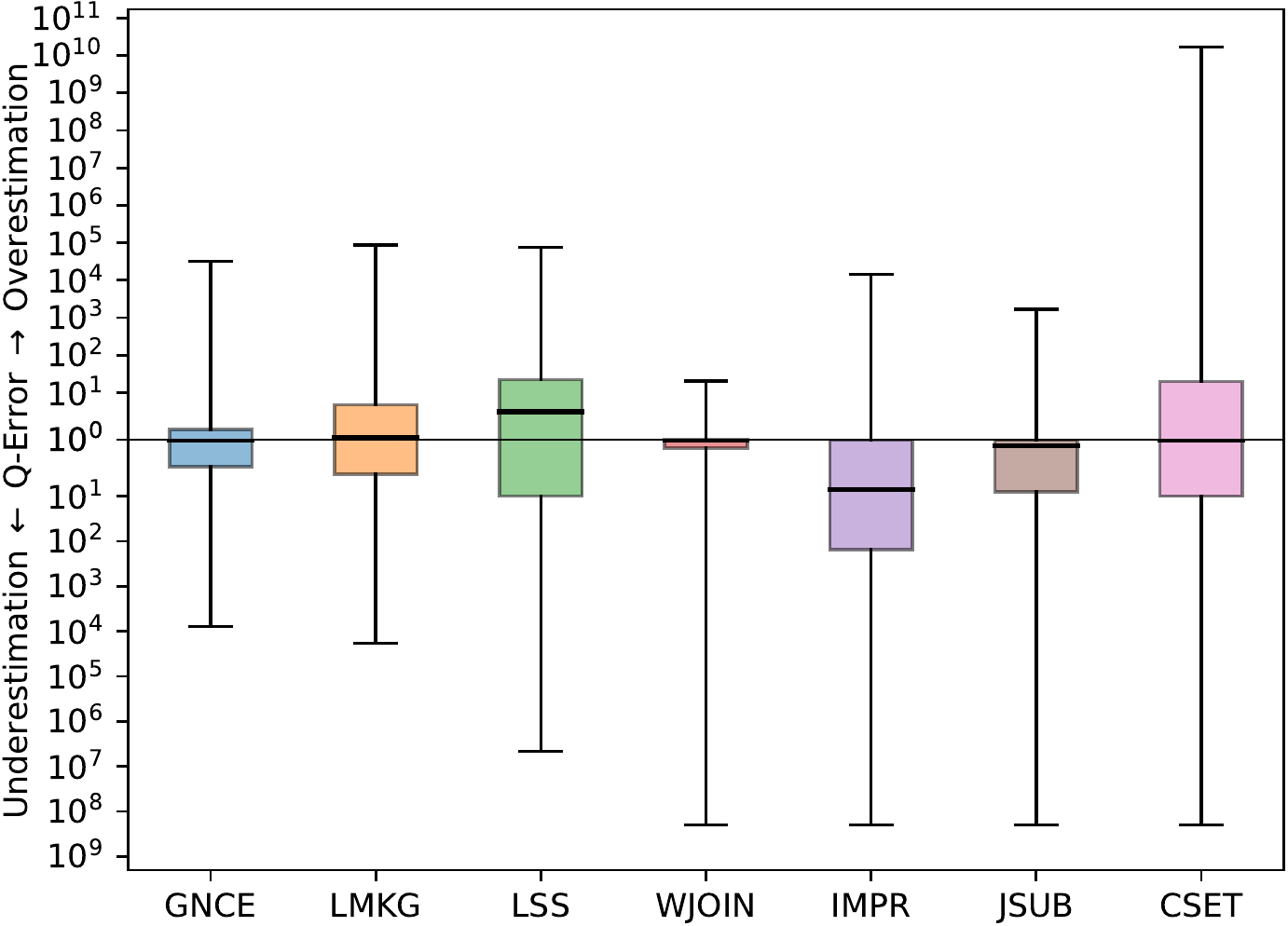}}
\caption{Results for complex queries on YAGO and Wikidata user queries}
\label{fig:plots_user}
\end{figure*}

Real user queries were retrieved from the Wikidata SPARQL query service log\footnote{\url{https://iccl.inf.tu-dresden.de/web/Wikidata_SPARQL_Logs/en}}.
We selected conjunctive queries that generated answers over our used Wikidata subset, yielding 11,832 queries, 90\% of which had a single triple pattern. We note that this high number might be due to filtering for conjunctive queries and ones that produce answers over the subset, as other works report a lower percentage of $\sim$77\% ~\cite{bonifati2020analytical}.
The learning-based methods were initially trained on the synthetic star and path queries, then finetuned on 80\% of the user queries. All methods are evaluated on the remaining 20\%. \textit{LSS} could only be evaluated on queries with more than one triple pattern because the provided code does not support single triple pattern queries.
Table \ref{q_table_user} shows mean q-Errors, where \textit{GNCE} outperforms all other methods. Figure \ref{fig:plots_user}(a) shows the mean q-Error, grouped by query sizes. \textit{LMKG} performs best in the single triple pattern case, while \textit{GNCE} performs best in the remaining ranges. The corresponding boxplot (Figure \ref{fig:plots_user}(b)) shows similar results to the synthetic queries.

\begin{table}[t!]
\caption{Mean q-Errors for complex and user queries}
\label{q_table_user}
\centering
\footnotesize
\begin{tabular}{|l|r|r|r|r|r|r|r|r|}
\hline
Query & \textit{Wanderjoin} & \textit{impr} & \textit{jsub} & \textit{CSET} & \textit{SUMRDF} & \textit{LMKG} & \textit{LSS} & \textit{GNCE} \\ \hline
  User &  3312 & 21984 & 3426 & $46 \cdot e^{11}$ & - & 268 & 198 & \textbf{13} \\ \hline
  Complex & 226282 & 508309 & 227163 & $3 \cdot e^{27}$ & - & 113 & 2519 & \textbf{8.4} \\ \hline
\end{tabular}%
\vspace{-2mm}
\end{table}
For more complex shape queries, we generated flower and snowflake queries over the YAGO dataset. 
A flower consists of a path query with one star at one end.  
A snowflake is a path, where at both ends a star is attached. 
We generated a total of 72,742 complex queries with 2 to 7 triple patterns. 
Table \ref{q_table_user} shows the mean q-Errors, where all methods perform worse compared to star and path queries over YAGO, although the distribution of query sizes is comparable. 
That shows that the complexity of the query structure directly impacts the performance of cardinality estimators. Further, \textit{GNCE} once more outperforms the other approaches in terms of the mean q-Error. Figure~\ref{fig:plots_user} (c) and (d) detail error distributions, reiterating \textit{Wanderjoin}'s good performance at low cardinality, and \textit{GNCE}'s dominance at high cardinality. We further observe that the boxplot of \textit{Wanderjoin} has the narrowest box around perfect predictions, i.e., it produces high-quality predictions for a large number of queries, accompanied however by large outliers (shown in the whiskers). This is consistent with the results by Park et al.~\cite{Park2020}.


\subsection{Inductive Case}
\label{sec:inductive_case}
We now test the effectiveness of the approaches in the inductive case, where test queries solely consist of entities not seen during training. Note that this is the worst-case scenario of inductivity, as typically some entities might have been seen before.
The inductive setting is of utmost importance for production environments, as it entails that the models can be trained using query samples containing a (proper) subset of the KG entities. 
On the contrary, models would require query samples that involve all the KG entities, which makes the deployment of such models rather unpractical.

We conducted this experiment on star queries for SWDF and YAGO and proceeded as follows. 
For each dataset, the query set was partitioned into two disjoint sets (a training- and a test), such that the entities in both sets were mutually exclusive, i.e., each query in one set did not involve any entity present in the other set (the predicates, however, were known, as is the usual setting for inductivity). Additionally, we removed all triples involving those entities from the input knowledge graph prior to building the index structures for summary- and sampling-based methods in \textit{G-CARE}. In the case of \textit{GNCE} and \textit{LSS}, no embeddings and occurrences were provided as featurization for the entities in the test set. Instead, a random (but fixed) vector was used to represent those entities. Those procedures ensured that the entities in the disjoint test set were never seen by any of the methods, including the statistics/sampling-builder, the embedding generator, or the neural networks. 

\begin{table}[]
  \caption{Mean q-Error for full inductive case on star queries}
  \label{fig:inductive_result_modified}
  \centering
  \footnotesize
  
    \begin{tabular}{|r|r|r|r|r|r|r|r|}
      \hline
      \multicolumn{8}{|c|}{Dataset: SWDF}\\
      \hline
      {$||Q||_\mathcal{G}$} & \textit{GNCE} & \textit{LMKG} & \textit{LSS} & \textit{Wanderjoin} & \textit{impr} & \textit{jsub} & \textit{CSET} \\ \hline
      \(< 5\) & 5.23 & 3.06 & \textbf{1.74} & 2.82 & 2.82 & 2.82 & 9.73 \\ \hline
      \(< 5^2\) & 3.16 & \textbf{2.88} & 4.26 & 12.51 & 12.51 & 12.51 & 11.68 \\ \hline
      \(< 5^3\) & \textbf{3.26} & 7.16 & 15.57 & 54.81 & 54.81 & 54.81 & 37.15 \\ \hline
      \(< 5^4\) & \textbf{4.16} & 29.92 & 89.02 & 264.14 & 274.87 & 284.91 & 176.37 \\ \hline
      \(< 5^5\) & \textbf{7.98} & 159.26 & 596.94 & 1008.72 & 1010.95 & 1011.22 & 538.91 \\ \hline
      \(< 5^6\) & \textbf{11.23} & 575.22 & 3567.94 & 1649.89 & 1655.85 & 1657.48 & 779.65 \\ \hline
      \(< 5^9\) & \textbf{335.51} & 840.27 & 65214.02 & 14198.33 & 14204.75 & 14198.37 & - \\ \hline
      Avg & \textbf{52.93} & 231.11 & 9927.07 & 2455.89 & 2459.51 & 2460.30 & 2250.30 \\ 
      \hline
      \multicolumn{8}{|c|}{Dataset: YAGO}\\
      \hline
      {$||Q||_\mathcal{G}$} &  \textit{GNCE} & \textit{LMKG} & \textit{LSS} & \textit{Wanderjoin} & \textit{impr} & \textit{jsub} & \textit{CSET}\\ \hline
      \(< 5\) & 189.48 & 319.94 & 8.58 & \textbf{2.51} & \textbf{2.51} & \textbf{2.51} & 12133.80 \\ \hline
      \(< 5^2\) & 229.71 & 138.35 & \textbf{3.94} & 11.60 & 13.55 & 13.55 & 13244.83 \\ \hline
      \(< 5^3\) & 73.98 & 47.19 & \textbf{8.10} & 33.96 & 50.27 & 42.74 & 25952.18 \\ \hline
      \(< 5^4\) & \textbf{38.18} & 145.65 & 41.76 & 185.62 & 262.63 & 218.33 & 13092.83 \\ \hline
      \(< 5^5\) & \textbf{44.89} & 595.18 & 196.23 & 628.17 & 1036.19 & 1012.65 & 64086.77 \\ \hline
      \(< 5^6\) & \textbf{86.65} & 2585.38 & 917.44 & 4249.70 & 6114.76 & 4277.01 & 119876.0 \\ \hline
      \(< 5^9\) & \textbf{280.94} & 49462.4 & 19016.3 & 141838.6 & 171508.1 & 142063.0 & 145802.6 \\ \hline
      Avg & \textbf{134.83} & 7613.44 & 2884.61 & 20992.88 & 25569.72 & 21089.98 & 56312.71 \\ \hline
    \end{tabular}
\end{table}

Then, the three learned approaches, \textit{GNCE, LMKG, LSS}, are trained on the training set (with embeddings for \textit{GNCE} and \textit{LSS}), and evaluated on the test set (without embeddings and occurrences). The methods from \textit{G-CARE} are evaluated on the test set using the statistics built with the KG excluding the entities in this test set.

The mean q-Errors are presented in Table~\ref{fig:inductive_result_modified}.
The results show that either \textit{GNCE, LMKG, LSS} or \textit{Wanderjoin} perform the best for specific query sizes. 
Compared to the non-inductive case in Figure~\ref{fig:bar_star_swdf}, all methods experience a performance deterioration, which is expected since statistics about the entities are missing. Overall, \textit{GNCE} performs the best.

In summary, these results show that \textit{GNCE} can be used in the inductive case involving new entities, without frequent retraining.

\subsection{Runtime Efficiency}
\label{sec:efficiency}

\subsubsection{Prediction Time (Online). }
We study the runtime of cardinality predictions for the different queries from the smallest and largest datasets, i.e., SWDF and YAGO.
To measure the runtime we use the \textit{time} library~\cite{python_time} in Python. 
For the learned techniques, we measure 1.) upon receiving a query the loading time of embeddings from disk and creation of the query featurization and 2.) the propagation of the query representation through the network's forward pass, which generates the final cardinality estimate. 
\begin{figure*}[h!]
\centering     
\subfigure[SWDF - Star]{\label{fig:time_swdf_star}\includegraphics[draft=false, width=0.45\textwidth]{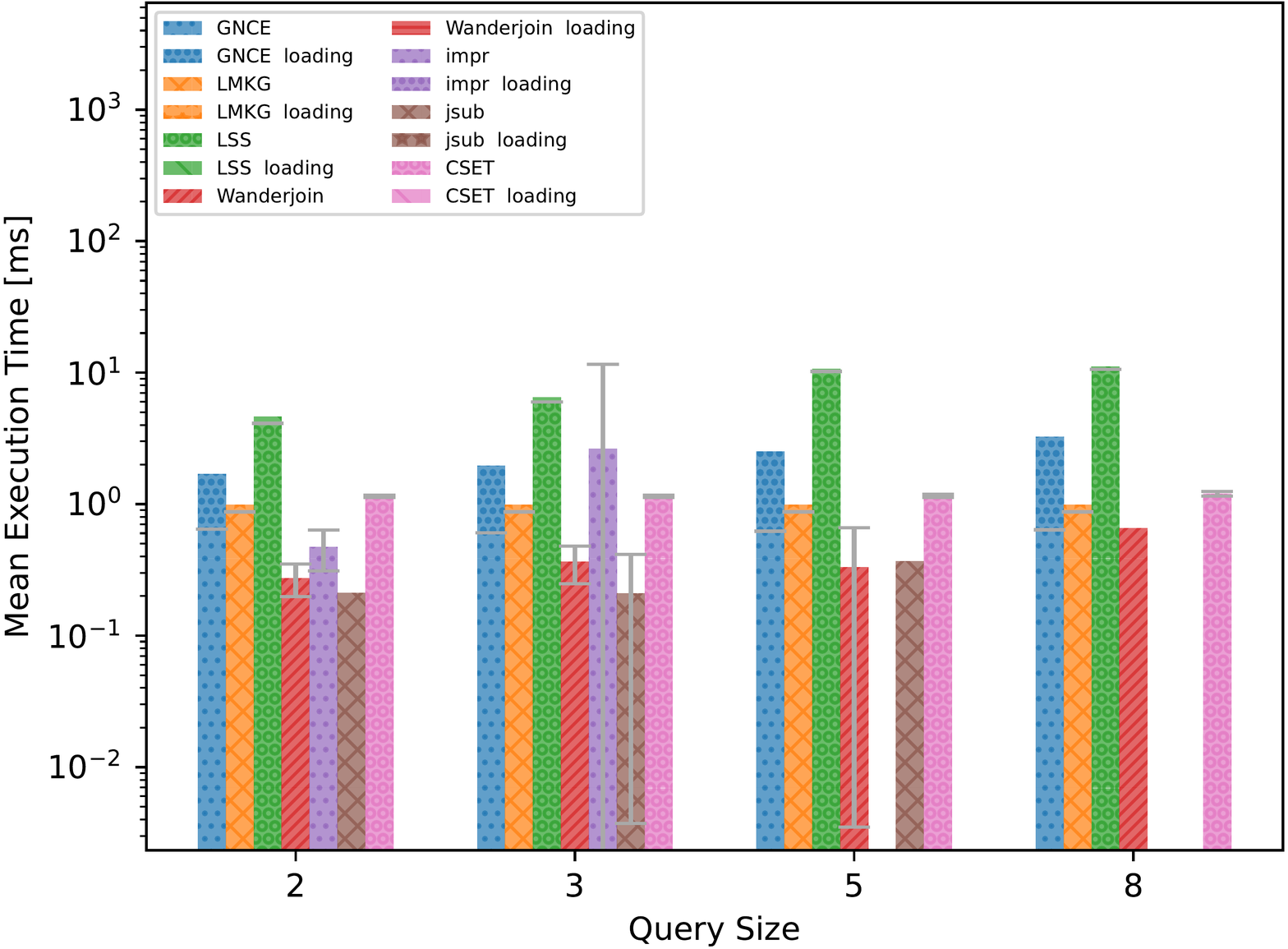}}
\hfill
\subfigure[SWDF - Path]{\label{fig:time_swdf_path}\includegraphics[draft=false, width=0.45\textwidth]{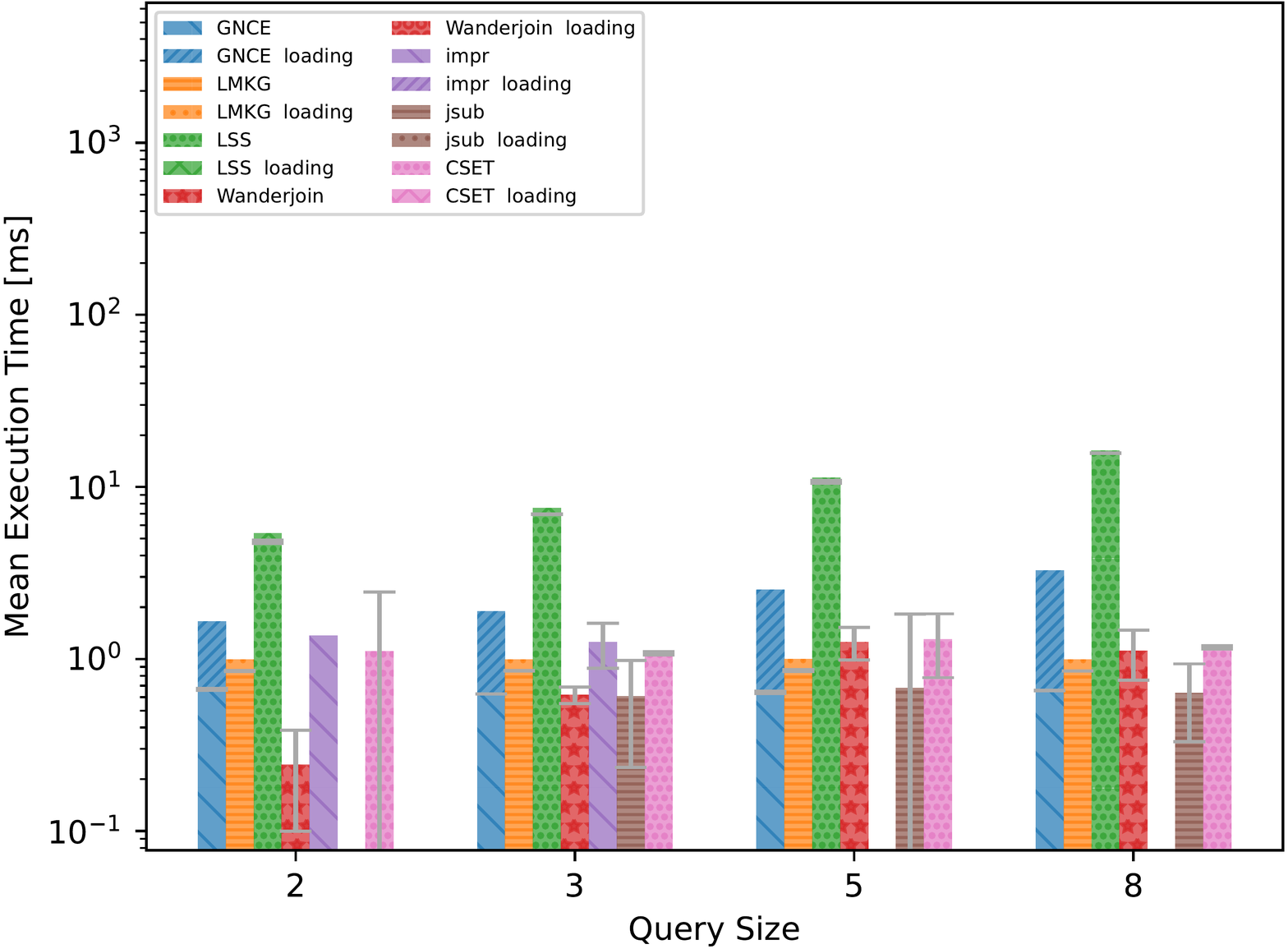}}

\subfigure[YAGO - Star]{\label{fig:time_yago_star}\includegraphics[draft=false, width=0.45\textwidth]{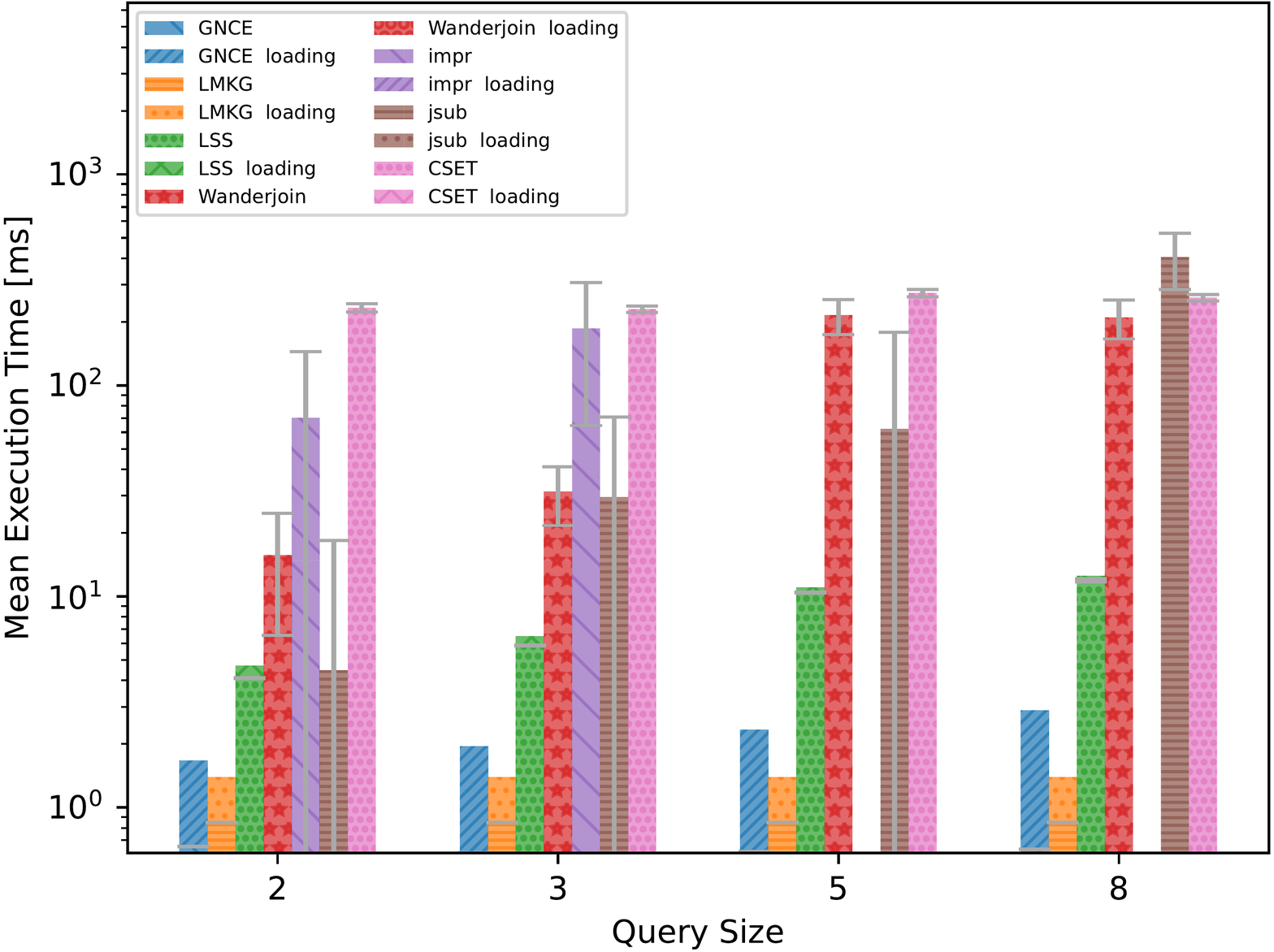}}
\hfill
\subfigure[YAGO - Path]{\label{fig:time_yago_path}\includegraphics[draft=false, width=0.45\textwidth]{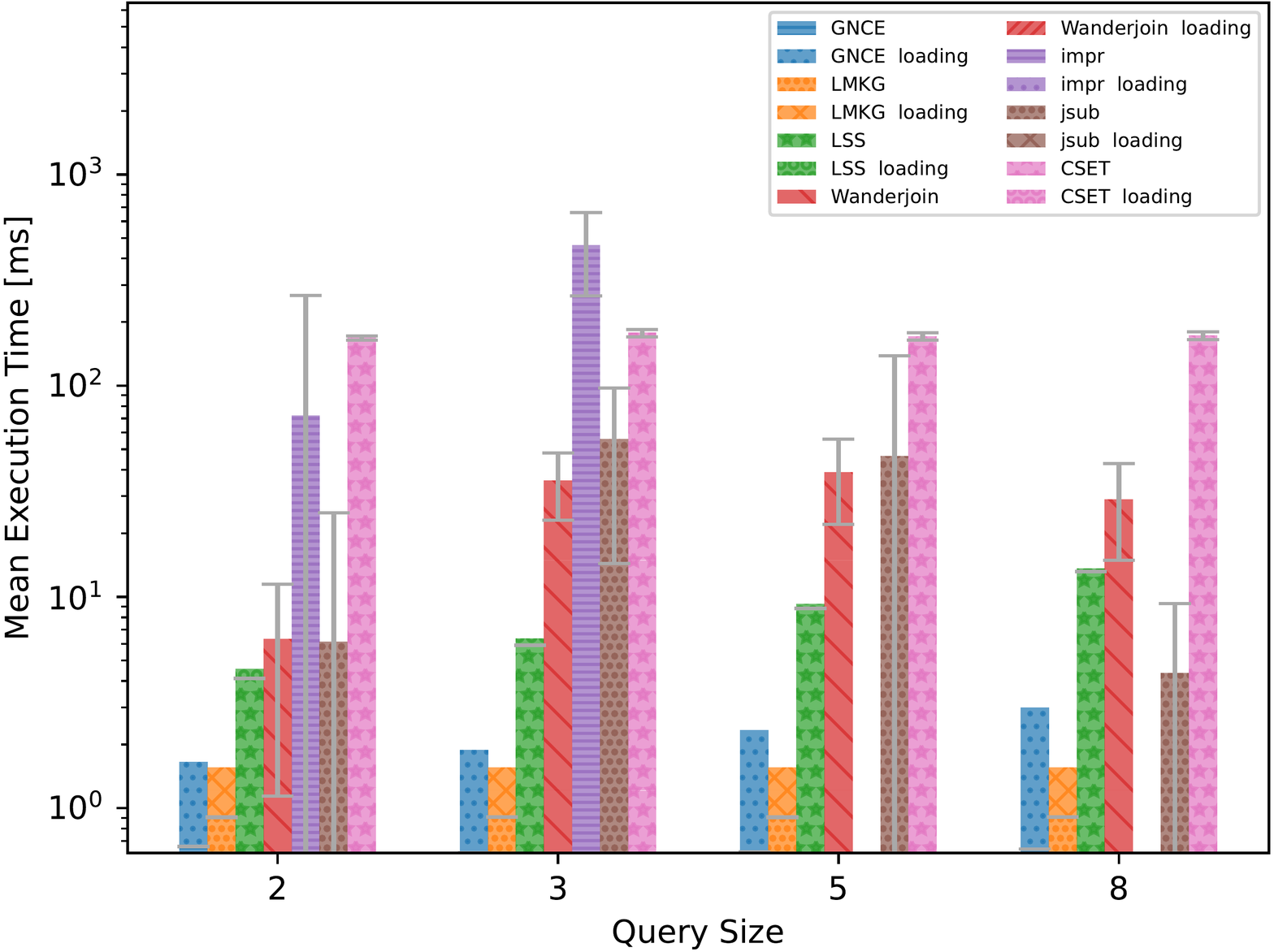}}
\caption{Mean prediction time [ms] of the different approaches, grouped by the number of triple patterns in the query. Confidence intervals (95\%) are shown in grey bars}
\label{fig:execution_plot}
\end{figure*}
For the methods provided by \textit{G-CARE}~\cite{Park2020}, we use the provided timing functionality of the different methods, which measures the total time from receiving a query to generating a cardinality estimate. For the sampling-based approaches, as per \textit{G-CARE} defaults, 30 samplings are performed, and their total runtime is used.
For all methods, we exclude failed cardinality predictions, i.e. cases where the cardinality estimate was $\leq 0$.

Figure~\ref{fig:execution_plot} presents the results grouped by the number of triple patterns in the query (\textit{query size}). 
The gray bars indicate the 95\% confidence interval. 
Comparing the results between SWDF and YAGO, we observe that the runtime of the non-learning-based approaches is impacted by the size of the dataset as well as the type of query. 
The sample-based methods, i.e. \textit{impr}, \textit{jsub}, and \textit{Wanderjoin}, show higher efficiency only on the smaller SWDF dataset.  
The performance of \textit{CSET} is similar to the sample-based approaches, as they require several accesses to (large) summaries to perform the predictions. 
We note that no runtimes for \textit{impr} are reported for query sizes 5 and 8, since all cardinality estimates failed for those sizes.
The model prediction times of the learned approaches exhibit stable behavior across the datasets, query sizes, and types. This is observed in both the mean values as well as the narrow confidence interval indicating almost no variations in runtime. The loading time for \textit{GNCE} and \textit{LSS} increases with the size of the query. This is expected since more embeddings need to be loaded from disk.
We observe that the GNN runtimes of \textit{GNCE} exhibit the best performance from the learned approaches and outperform the sampling/statistics-based methods. We further note that the majority of the total runtime for \textit{GNCE} is due to embedding loading and query graph featurization, but overall still runs within 1-2 ms. Using efficient storage and compiled code can reduce \textit{GNCE} runtime significantly. Overall, \textit{LMKG} displays the best runtime of the methods.
On the larger dataset, the runtimes of the learned methods are several orders of magnitude faster than the summary/statistics-based methods. 
Fast runtime within the single-digit millisecond range is a desired result since a very fast runtime is essential for a cardinality estimator to run, for example, within a query optimizer.

\subsubsection{Building Model/Statistics Time (Offline).}
All the studied methods need to build/train a model or build statistics/samples prior to being used as a cardinality estimator. Hence, we now measure the time it takes to do this preparation. We evaluate this using YAGO, but present building/training (refer in the remainder as training) times per atom, in order to easily interpolate between methods and to other graphs. 
The learned methods are evaluated with a Batch Size of 1, and a total of 20 training epochs.

\begin{table}[h!]
    \caption{Embedding and training times per atom (in ms)}
    \centering
    \label{tab:training_time}
    \footnotesize
    \begin{tabular}{|l|c|c|c|c|c|c|c|}
        \hline
        & \textit{GNCE} & \textit{LMKG} & \textit{LSS} & \textit{Wanderjoin} & \textit{impr} & \textit{jsub} & \textit{CSET} \\
        \hline
        Embedding & 6.40 & - & 3.94 & - & - & - & - \\
        \hline
        Training & 2.52 & 1.16 & 114.03 & 0.00019 & 0.00020 & 0.00019 & 0.00022 \\
        \hline
    \end{tabular}
\end{table}


Table~\ref{tab:training_time} shows the results of this study. 
Comparing the different values for the runtimes in the test case and the training times, we can see that the training times for sampling/summary-based methods are significantly faster than the learning-based methods, while the test times are faster for the learning methods (on larger graphs).
The intuitive explanation for this is that during NN training, lots of statistics and patterns about the KG are compressed into the embeddings and NN parameters, which takes longer but enables a faster execution time.
The sampling/summary-based methods compute less statistics up front but need to perform more computation during prediction, especially if the KG is large. Longer training times for the learned approaches prior to usage can be tolerated, but a very fast execution time (especially in the case of query optimization) is required.
We also want to mention that the significantly largest time for \textit{GNCE} is due to the embedding calculation, which was performed on a relatively small machine. 
During embedding learning, most of the time is spent on random walk generation, a process that can be parallelized if the hardware allows. To put this into perspective, with the used hardware, calculating embeddings for 7 million entities (approximate size of DBpedia~\cite{lod-cloud-dbpedia}) would take $\sim$12 hours. Then, training the \textit{GNCE} GNN with 100,000 queries with an average of 5 triple patterns would take $\sim$ 63 minutes.
\subsection{Ablation Study}
\label{subsec:ablation}
To determine the impact of the different components of \textit{GNCE}, we performed an ablation study. 
For that, we compared the baseline model (\textit{GNCE}) that was used in the previous sections to the same model where (1) the RDF2Vec embeddings are replaced by binary encodings, and (2) the directional TPN message passing is reduced to an undirected message passing where no case distinction between incoming and outgoing edges is performed. 
For (1), all entities in the KG are enumerated and given an \textsc{id}, and the corresponding embedding of the entity is replaced by a binary representation of the \textsc{id}, padded with 0s to length $100$. 
For the sake of space, the study is performed only on the largest KG YAGO.

Table \ref{tab:ablation} shows that removing RDF2Vec embeddings from the query graphs deteriorates ($\downarrow$) performance for both star and path queries. 
However, the effect is much stronger for path queries. 
This can be explained by the fact that the predicates of the star are already very predictive of the query cardinality (similar to characteristic sets). For the same reason, removing directed message passing also only deteriorates performance on path queries.

\begin{table}[t!]
\caption{Ablation study results of \textit{GNCE} on YAGO}
\label{tab:ablation}
\centering
\begin{tabular}{l|l|l}
Method                    & Star Queries & Path Queries \\ \hline
Baseline                  &     1.96         &    3.78          \\ \hline
-RDF2Vec                  &         2.40 ($\downarrow$ 22\%)     &     6.14 ($\downarrow$ 62\%)         \\
-Directed Message Passing &        1.93 ($\uparrow$ 1\%)      &          3.87 ($\downarrow$ 3\%)   
\end{tabular}
\end{table}

\subsection{Application of GNCE to Query Optimization}
\label{sec:optimization}
As mentioned, the overarching goal of cardinality estimation is to provide estimates to the query optimizer to minimize the query runtimes. 
Even though query optimization is an orthogonal problem to the one tackled in this work, we demonstrate preliminary results of \textit{GNCE} assisting an actual query optimizer. 
For this, we have chosen nLDE~\cite{nlde}, an open-source SPARQL engine also implemented in Python. 
nLDE uses cardinality estimates of subqueries to select physical join operators while traversing the space of plans with greedy heuristics. 
We compared the query plans devised with the nLDE cardinality estimator vs. \textit{GNCE}.  

\begin{figure}[h!]
\subfigure[Star Queries]{\label{fig:b}\includegraphics[width=0.45\textwidth]{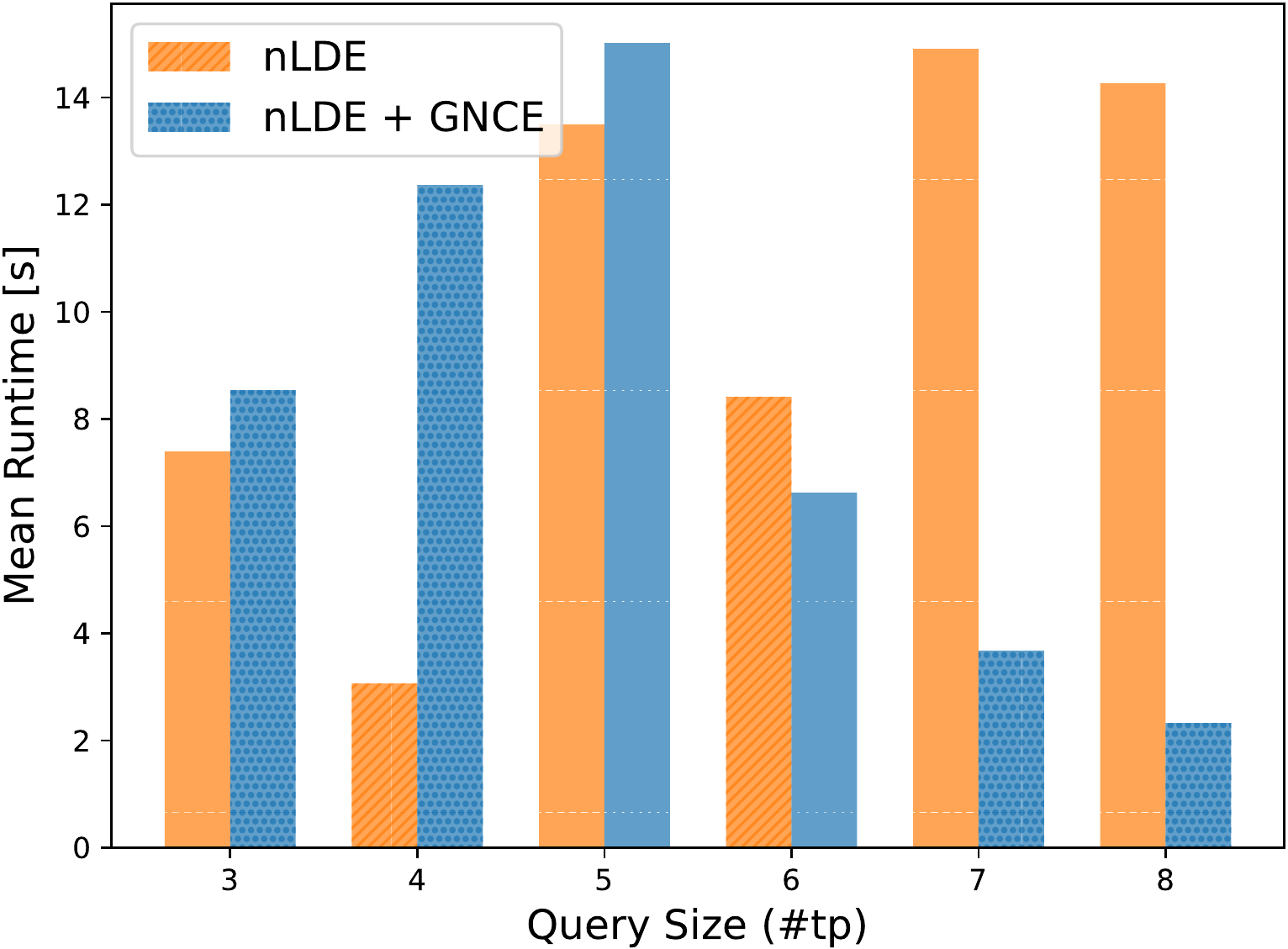}}
\hfill
\subfigure[Path Queries]{\label{fig:a}\includegraphics[width=0.45\textwidth]{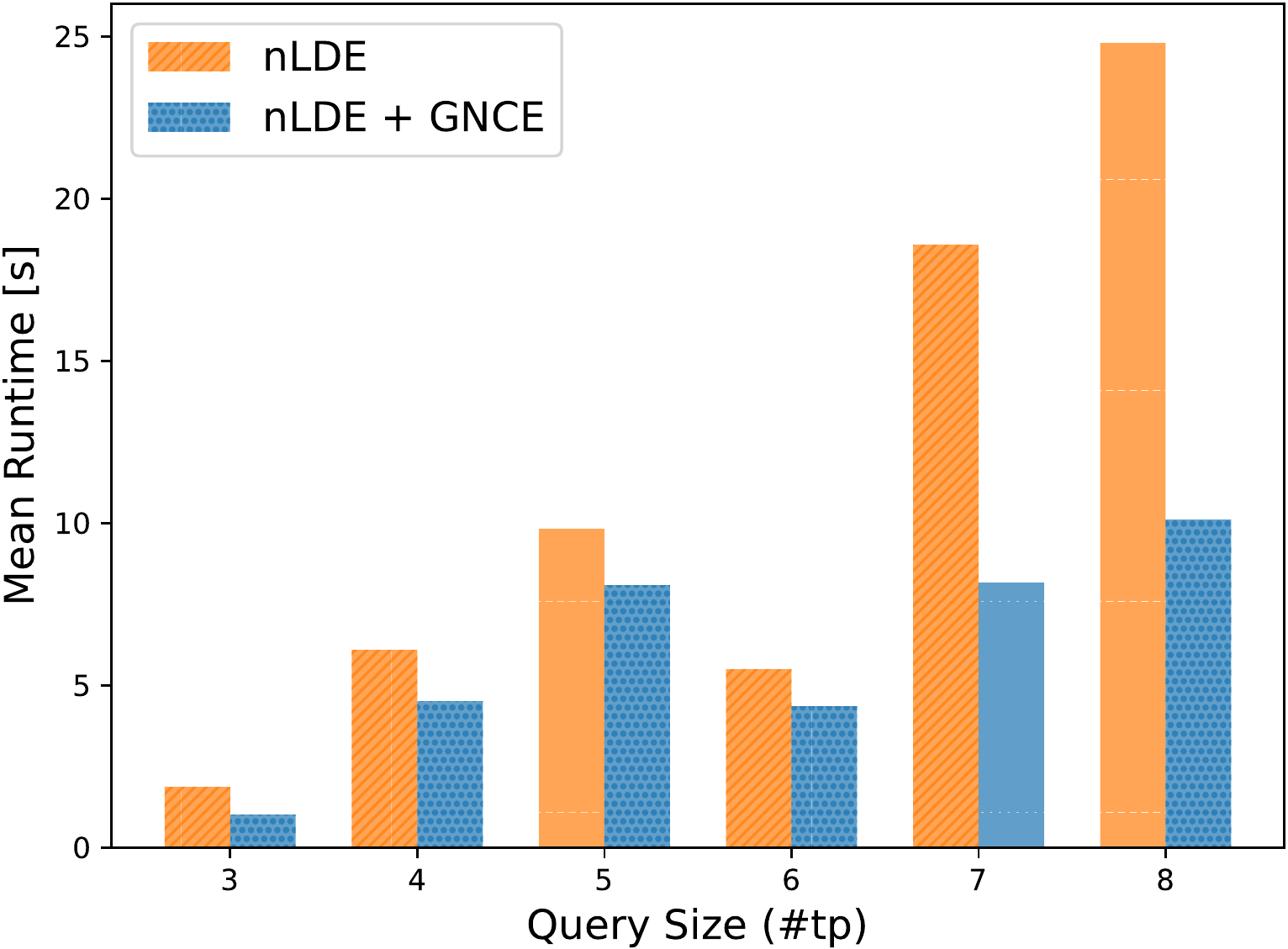}}
\caption{Query optimization results on SWDF}
\label{fig:query_optimization}
\end{figure}

For this study, we extended our SWDF query set to include queries with 3-8 triple patterns.
We tested 243 star queries and 258 path queries. Therein, 15\% of the star queries and 14\% of the path queries had different plans when using \textit{GNCE} as an estimator. Queries with the same plan under both cardinality estimators had similar runtimes, as expected.
Figure \ref{fig:query_optimization} shows the runtimes of star and path queries with different resulting plans.
In general, the larger the query size, the faster the query execution using \textit{GNCE}. For small star queries, \textit{GNCE} does not improve the execution times. For large queries (7 and 8 triple patterns), the execution time is improved by more than 100\%. 
As can be read from Figure~\ref{fig:time_swdf_star}, for a query with 8 triple patterns (i.e., 7 joins), it takes \textit{GNCE} 14ms to produce the cardinality predictions during optimization. This is approximately 0.1\% of the total execution time.
\section{Conclusion and Outlook}
\label{sec:conclusion}
We presented \textit{GNCE}, a method for accurate cardinality estimation of conjunctive queries over Knowledge Graphs (KGs) based on KG embeddings (KGE) and Graph Neural Networks (GNN). 
KGE provide semantically meaningful features to the GNN, which directly predicts the cardinality of the embedding-enhanced query graph. 

Our experiments show that \textit{GNCE} on average outperforms the state of the art in terms of q-Error over synthetic and real-world KGs and queries. 
\textit{GNCE} is also robust in the sense that it does not produce severe prediction errors. 
\textit{GNCE} also generalizes well to unseen entities, making it a suitable solution for actual deployment. 
In that context, we also showed that our model is lightweight 
leading also to smaller prediction runtime below 1~ms.

In future work, we plan to integrate queries with FILTERs.
This requires extensions on the query sampling and KGE random walks for better capturing underlying range distributions, and the query featurization to represent the filter operation and the value.
Further, we will investigate special treatments of numerical and text literals with more meaningful embeddings.
Incorporating active learning, by means of an adaptive sampling strategy with respect to poorly performing queries, with the goal of accelerating the GNN training, is another promising direction.
Lastly, a custom KGE approach tailored to cardinality estimation will be investigated.

\section{Acknowledgements}
The authors would like to thank Joscha Klimpel and Prof. Dr. Hanno Gottschalk for their valuable input in the preparation of the paper.

\bibliographystyle{unsrtnat}
\bibliography{references}  






\end{document}